\documentstyle[12pt,aaspp4]{article} 

\newcommand{\bq}{\begin{equation}} 
\newcommand{\eq}{\end{equation}}   
 
\newcommand{\ho}{km s$^{-1}$ Mpc$^{-1}$}

\newcommand{\PL}{$P$-$L$}

\begin{document} 
\def\refitem{\par\parskip 0pt\noindent\hangindent 20pt} 
\def\Sec{${}^{\prime\prime}$\llap{.}}

\normalsize 

\title{Cepheid Calibrations from the Hubble Space Telescope of the Luminosity
of Two Recent Type Ia Supernovae and a Re-determination of the Hubble
Constant\altaffilmark{1}}

\vspace*{0.3cm}  
{\it  \ \ \ \ \ \ \ \ \ Accepted; To Appear in the Astrophysical Journal}

\vspace*{0.3cm}
\noindent
Adam G. Riess\altaffilmark{2}, 
Weidong Li\altaffilmark{3},
Peter B. Stetson\altaffilmark{4},
Alexei V. Filippenko\altaffilmark{3},
Saurabh Jha\altaffilmark{3},
Robert P. Kirshner\altaffilmark{5},
Peter M. Challis\altaffilmark{5},
Peter M. Garnavich\altaffilmark{6}, and
Ryan Chornock\altaffilmark{3}

\altaffiltext{1}{Based on observations with the NASA/ESA {\it Hubble Space 
Telescope}, obtained at the Space Telescope Science Institute, which is 
operated by AURA, Inc., under NASA contract NAS 5-26555.} 
\altaffiltext{2}{Space Telescope Science Institute, 3700 San Martin 
Drive, Baltimore, MD 21218; ariess@stsci.edu} 
\altaffiltext{3}{Department of Astronomy, 601 Campbell Hall, University of 
California, Berkeley, CA  94720-3411; (weidong, alex, sjha, rchornock)@astro.berkeley.edu} 
\altaffiltext{4}{Herzberg Institute of Astrophysics, National
Research Council, 5071 West Saanich Road, Victoria, BC V9E2E7, Canada; 
Peter.Stetson@nrc-cnrc.gc.ca}
\altaffiltext{5}{Harvard-Smithsonian Center for Astrophysics, 60 Garden St., 
Cambridge, MA 02138; kirshner@cfa.harvard.edu}
\altaffiltext{6}{University of Notre Dame, Department of Physics, 225
  Nieuwland Science Hall, Notre Dame, IN 46556-5670; pgarnavi@miranda.phys.nd.edu}

\begin{abstract}

We report observations of two nearby Type Ia supernovae (SNe~Ia) for which
observations of Cepheid variables in the host galaxies have been obtained with
the {\it Hubble Space Telescope}: SN 1994ae in NGC 3370 and SN 1998aq in NGC
3982.  For NCG 3370, we used the Advanced Camera for Surveys to observe 64
Cepheids that yield a distance of 29 Mpc, the farthest direct measurement of
Cepheids.  We have measured emission lines from H~II regions in both host
galaxies which provide metallicity-dependent corrections to their
period-luminosity relations.  These two SNe~Ia double the sample of ``ideal''
luminosity calibrators: objects with well-observed and well-calibrated light
curves of typical shape and with low reddening.  By comparing them to all
similarly well-measured SNe Ia in the Hubble flow, we find that $H_0 = 73 \pm
4$ (statistical) $\pm 5$ (systematic) \ho.  A detailed analysis demonstrates
that most of the past disagreement over the value of $H_0$ as determined from
SNe~Ia is abated by the replacement of past, problematic data by more accurate
and precise, modern data.

\end{abstract} 
subject headings: galaxies: distances and redshifts ---
cosmology: observations --- cosmology: distance scale --- 
supernovae: general

\section{Introduction}

  Type Ia supernovae (SNe~Ia) are the most precise distance indicators known
for measuring the expansion rate of the Universe.  Their extreme and relatively
uniform luminosities provide the means to measure accurate and precise
distances across a significant fraction of the visible Universe (see Leibundgut
2001 for a review) and gave the first evidence of its accelerating expansion
(Riess et al. 1998; Perlmutter et al. 1999).  Methods which utilize the
relationship between SN~Ia light-curve shape and luminosity (Phillips 1993;
Hamuy et al. 1995, 1996a; Riess, Press, \& Kirshner 1995, 1996a; Perlmutter et
al. 1997) and supernova color to measure absorption by dust (Riess et
al. 1996b; Phillips et al. 1999) yield distances with {\it relative} precision
approaching 7\% {\it when applied to modern photometry}.  The tightest Hubble
diagram is provided by SNe~Ia (see Figure 1).  The $\sim$90 high-quality SN~Ia
light curves currently published in the range $0.01 < z < 0.1$ establish the
{\it relative} expansion rate to an unprecedented uncertainty of less than 1\%.
One-quarter of these SNe~Ia are from the Cal\'{a}n/Tololo survey (Hamuy et
al. 1996b) and the rest are from the CfA Survey I (Riess et al. 1999a) and the
CfA Survey II (Jha et al. 2005a).

   A calibration of the peak luminosity of SNe~Ia, while not necessary for past
measurements of {\it changes} in the expansion rate, is needed to fix the
distance scale of the Universe by determining the Hubble constant.  Many other
quantities of interest in cosmology and astrophysics directly depend on the
value of the Hubble constant, including the expansion age of the Universe, the
size of the visible Universe, the closure density, and the baryon density, to
name a few.  An accurate determination of $H_0$ also enables the extraction of
the mass density and energy density from the measurement of the distance to the
last-scattering surface, for example by the Wilkinson Microwave Anisotropy
Probe (Bennett et al. 2003; Spergel et al.  2003).

   An accurate and precise measurement of the Hubble constant was one of the
original motivations for the construction of the {\it Hubble Space Telescope
(HST)}, and two teams ambitiously tackled the challenge of determining $H_0$ by
analyzing {\it HST} observations of Cepheid variables (beginning even before the
replacement of WF/PC with WFPC2).
  
  The Type Ia Supernova {\it HST} Calibration Program, led by A. Saha, G. Tammann, 
and A. Sandage (hereafter STS; see Saha et al. 1997; Saha et al. 2001, and 
references therein) has produced calibrations of the luminosity of eight individual
SNe~Ia using $HST$ observations of Cepheid variables in their host galaxies.
However, the limited volume within which $HST$ with WFPC2 can resolve Cepheids
($R \approx 20$ Mpc) severely limits the quality of supernova data to calibrate.
Half of the SNe~Ia calibrated by STS were discovered before the advent of CCDs
(see Table 1).  The light curves of SN 1895B, SN 1937C, SN 1960F, and SN 1974G
were obtained with photographic emulsions and show {\it internal} dispersions
of $\sim$0.2 mag.  An accurate accounting of the background light is difficult
to achieve and frequently unreliable.  Inaccurate subtraction of background
light from such data can result in systematic errors in peak brightness and
light-curve shape, as described by Boisseau \& Wheeler (1991).  In addition,
the nonstellar spectrum of SNe~Ia makes it difficult to accurately convert
photographic magnitudes to standard passbands.

   Dramatic progress in the cosmological utility of SNe~Ia has been made
possible by improvements in the quality of their photometric data in the last
two decades.  It is time for the pursuit of Hubble's constant via SNe~Ia to
overcome the problems inherent in pre-digital data and benefit from superior
calibrations.

   Additional liens against past calibrators include objects with high
reddening leading to large and thus more uncertain extinction (e.g., $A_V \geq
1$ mag for SN 1989B and SN 1998bu) and objects with atypical light-curve shapes
leading to relatively large and thus more method-dependent luminosity
corrections (e.g., SN 1991T and SN 1999by).  The use of such objects risks
contaminating the fiducial SN~Ia luminosity determined from the small sample of
calibrators.  (This danger is not an issue for the Hubble-flow set of SNe~Ia 
which is large enough to overcome a few inaccurate objects.)

   Only two of the eleven calibrated SNe~Ia (SN 1981B and SN 1990N) are ideal.
These have been observed before maximum light, through low interstellar
extinction ($A_V < 0.5$ mag), with the same instruments (CCDs and the
Johnson/Cousins passbands) as the SNe~Ia in the Hubble flow from the
Cal\'{a}n/Tololo and CfA Surveys, and have typical light-curve shapes.  
Interestingly, the average magnitude of these two SNe~Ia is 0.14 mag fainter
than the average of the other 7 less optimal calibrators (Saha et al. 2001).
Clearly the best course of action will be to increase the small sample of such
calibrators.

   Another impediment to an accurate resolution of the Hubble constant has been
disagreement over the analysis of Cepheid data.  A subset of the $HST$ Key
Project (hereafter referred to as ``SKP'') has reached significantly different
conclusions about the value of $H_0$ from precisely the same data utilized by
the STS group (Gibson et al. 2000).  By reanalyzing the Cepheid data of STS,
the SKP concludes that each of the SN~Ia calibrators is intrinsically 0.2--0.3 mag
fainter (even when boths groups utilize the same 18.50 mag distance modulus of
the Large Magellanic Cloud (LMC) and agree on the measured fluxes of the $HST$
Cepheids).  We believe that an independent look at these differences is warranted.
We find that the origin of at least some of the differences in interpretation
rests in pernicious anomalies in WFPC2 which can now be circumvented by the new
Advanced Camera for Surveys (ACS) on $HST$.

   The other improvement offered by ACS is that $HST$ can now well-measure the
light curves of Cepheids in an economical amount of observing time at a
distance of up to $\sim$ 30 Mpc.  This increased range provides an opportunity
to add to the small calibrated sample of ideal SNe~Ia.  In \S 2 we present
observations of Cepheid variables in NGC 3370 obtained with ACS on $HST$ in a
55-day campaign during Cycle 11.  NGC 3370 was the host of SN 1994ae, a
spectroscopically normal (Filippenko 1997; Branch, Fisher, \& Nugent 1993)
SN~Ia observed through $UBVRI$ filters with modern CCDs beginning 12 days
before maximum brightness (Riess et al. 1999b).  The SN has $\lesssim 0.25$ mag
of visual extinction from its host, reducing uncertainties from unknown
extinction laws.  In \S 3 we present a careful recalibration of the light curve
using additional photometric calibrations of the field, and galaxy template
subtraction to remove contamination of the SN photometry.  We also present the
first photometric record of SN 1998aq, a similarly well-observed prototypical
SN~Ia with low reddening.  Combined with the recent Cepheid calibration of its
host, NGC 3982 (Saha et al. 2001; Stetson \& Gibson 2000), we calibrate its
luminosity.  These two SNe~Ia double the sample of ideal SNe~Ia whose
properties and photometric records can provide the most reliable absolute
calibrations for the SN~Ia Hubble diagram.  In \S 4 we present estimates of
$H_0$, and we discuss the analysis in \S 5.

\clearpage

\begin{table}[h]
\begin{small}
\begin{center}
\vspace{0.4cm}
\begin{tabular}{cccccc}
\multicolumn{6}{c}{Table 1: SN~Ia Light Curves with $HST$ Cepheid Calibration} \\
\hline
\hline
SN Ia & CCD photometry? & low reddening?$^a$ & observed before maximum? & normal & ideal? \\
\hline
1895B$^b$  & No & Unknown & No & Unknown & \\
1937C$^b$ & No & Yes & Yes & Yes & \\
1960F$^b$ & No & No & Yes & ? & \\
1972E$^b$ & No & Yes & No & Yes & \\
1974G & No & No & Yes & ? & \\
1981B$^b$ & Yes & Yes & Yes & Yes & $\surd$ \\
1989B$^b$ & Yes & No & Yes & Yes & \\
1990N$^b$ & Yes & Yes & Yes & Yes & $\surd$ \\
1991T$^b$ & Yes & Yes? & Yes & No & \\
1994ae$^c$ & Yes & Yes & Yes & Yes & $\surd$ \\
1998aq$^c$ & Yes & Yes & Yes & Yes & $\surd$ \\  
1998bu & Yes & No & Yes & Yes & \\
1999by & Yes & Yes & Yes & No & \\

\hline
 \hline
\multicolumn{6}{l}{$^aA_V < 0.5$ mag.} \\
\multicolumn{6}{l}{$^b$Calibrated by the STS collaboration.} \\
\multicolumn{6}{l}{$^c$Calibration first presented in this paper.} \\
\end{tabular}
\end{center}
\end{small}
\end{table}

\clearpage

\section{The Cepheid-Calibrated Distance to NGC 3370}

  \subsection{Observations}

   NGC$\,$3370 is a relatively small (3.0$^\prime \times 2.5^\prime$), nearly
face-on Sc galaxy (see Figure 2), identified as a member of a loose group (the
Lyon Group LGG 219; Garcia 1993) which includes four other spirals and one
irregular.  Its heliocentric recession velocity is 1280 km s$^{-1}$ as measured
from systemic 21 cm H~I (de Vaucouleurs et al. 1991).  Based on the peak
brightness of SN 1994ae ($V = 13.1$ mag; Riess et al. 1999b), the low reddening
at maximum ($B-V = 0.1$ mag), and previous calibrations of the peak luminosity
($M_V = -19.4$ mag; Jha et al. 1999; Saha et al. 2001), we anticipated a
distance modulus of $\sim$32.4 mag (distance about 30 Mpc) when proposing for
$HST$ observations of Cepheids in this host.

   Given the late-type morphology, low inclination, and estimated distance,
NGC$\,$3370 appeared to be an excellent candidate for hosting Cepheid variables
whose light curves could be measured in an economical amount of time with ACS.
We used exposure times of 2 orbits (4800~s) per epoch, which was expected to
give a signal-to-noise ratio (S/N) of 10--25 for Cepheids of $20 < P < 60$ days
($27.4 < F555W < 26.1$ mag), similar to what was obtained by the SKP and STS in
most of their programs.  In all, we observed for twelve epochs in $F555W$ with
power-law interval spacing over a 53-day span to reduce aliasing (Freedman et
al. 1994) and five epochs in $F814W$ to aid in the identification of Cepheids
and to measure the mean dust extinction from their colors.  The exposure 
information is contained in Table 2.  

The telescope orientation and position
were fixed throughout the observations to simplify the data reduction (but
small shifts of 1--2 pixels are routine for visits which use the same guide
stars throughout).  Imaging at each epoch was obtained at four dither positions
(each of 1200~s duration): two positions with a small (non-integer) shift of a
few pixels in the $x$ and $y$ directions repeated after a large shift of 60
pixels in the $y$ direction to cover the ACS chip gap.  (The reason for filling
in the gap was to produce an aesthetically pleasing image suitable for a Hubble
Heritage Program press release.)  The orientation and position were chosen to
contain a few external (to NGC$\,$3370) bright sources in the ACS field which
were also visible in the original monitoring data of SN 1994ae (Riess et
al. 1999b) from the 1.2-m telescope at the Fred Lawrence Whipple Observatory
(FLWO), as seen in Figure 2. (A similar view of SN 1998aq and $HST$ images of
its host are shown in Figure 3.)  Matching these sources provides the means to
register the $HST$ images to the ground-based SN follow-up data and allow us to
search for a possible light echo or late-time visibility of SN 1994ae with high
astrometric precision (though ultimately neither was detected to the limit of the
frames, $F555W \approx 29.0 $ mag).

   After acquisition, all images were processed using up-to-date reference
files and the CALACS pipeline in the STSDAS package in IRAF\footnote{IRAF is
distributed by the National Optical Astronomy Observatories, which are operated
by the Association of Universities for Research in Astronomy, Inc., under
cooperative agreement with the National Science Foundation.}. This procedure
includes ``standard'' rectifications for the camera gain, overscan, spatial
bias, dark current, and flat fielding.  Due to the significant geometric
distortion of the ACS WFC (the cost of minimizing reflections), we applied the
``drizzle'' algorithm (Fruchter \& Hook 1997) in the Multidrizzle software
package (Koekemoer et al. 2005).  Because ACS WFC images are undersampled at
wavelengths shortward of 11,000~\AA, a better sampled and more precise
point-spread function (PSF) can be obtained by ``drizzling'' (i.e., resampling
and combining) the images at a pixel scale finer than the physical detector
pixels of 0\Sec05 pixel${}^{-1}$.  However, full resampling can only be
realized with well-dithered images.  The relative size of the dither was
measured for each frame using source catalogs.  The images were subsequently
resampled to 0\Sec033 pixel${}^{-1}$.

   ACS PSFs were constructed in the usual way from uncrowded frames of
47~Tucanae (NGC$\,$104) obtained by an ACS calibration program (GO
9648).  Since only one (drizzled) image of 47~Tuc was used to define
the PSF for each filter, the normal DAOPHOT ``PSF'' command was used,
rather than the more powerful program MULTIPSF (Stetson 1993).
Approximately 250 stars well-distributed across the 47~Tuc field were
used to define PSFs with quadratic spatial variation for the three
filters.  The derived PSFs did a good job of subtracting the stars
from the images, so we infer that the spatial variations in the
(geometrically rectified) images are well-represented by our model.

   Since the drizzled images employed here were already largely free of cosmic-ray
events, we employed a reduction process similar to that used for reducing
ground-based data.  Catalogs of stellar-appearing objects in 47~Tuc and NGC$\,$3370
were produced by the ``FIND'' command in DAOPHOT, and initial magnitude
estimates and local sky-brightness values were obtained with the ``PHOT''
command.  At this point the aforementioned PSFs were obtained from the images
of 47~Tuc by an iterated application of the ``PSF'' command and the program
ALLSTAR (see, e.g., Stetson 1987).  These PSFs were then employed
to analyze the NGC$\,$3370 images with ALLSTAR to produce improved magnitude and
sky-brightness estimates for the detections in the individual images.  This
process normally results in the rejection of some poor-quality and evidently
nonstellar detections.  These refined individual star lists were then
intercompared and merged by the programs DAOMATCH and DAOMASTER, a process which
produces both a master list of unresolved sources in the field and a set of geometric
transformations relating the coordinate systems of the individual images to a
common reference frame.  These are the raw inputs to the program ALLFRAME, which
enforces self-consistent positions for the detected objects as it produces
improved estimates of the individual-epoch magnitudes and geometric
transformations.  Then the original and star-subtracted images were stacked and
examined by eye to identify star-like images that had been missed by the
automatic routines.  These were added to the master list for the field, and the
DAOMASTER and ALLFRAME reductions were repeated until we judged that all the
stellar objects that we could see had been adequately reduced.

   Photometric calibration of the sources was derived in two independent
ways. One method, similar to what was done by the SKP, was to derive
photometric transformations by matching the ACS data for 47~Tuc in $F435W,$
$F555W,$ and $F814W$ to ground-based photometry for stars in the cluster.  This
photometry was derived from the same images employed by Stetson (2000) to define
secondary standard stars in the cluster field.  These observations were obtained on
22 nights from 12 distinct observing runs with various telescopes (Stetson 2000).  Calibrated
magnitudes were extracted from the master data set for the same 250 hand-selected
stars as were used to define the PSFs in the ACS images of the 47~Tuc field.
The median number of independent measurements of the magnitude of each star was
31 in each of $B$ and $V$ (minimum 25, maximum 32), and 36 (30, 37) in $I$,
and the median photometric uncertainty of the derived ground-based magnitudes
was 0.003$\,$mag in $B$, 0.002 in $V$, and 0.002 in $I$.
A ground-based color-magnitude diagram for these 250 local standards is shown
as Figure 4.  While there is only a sprinkling of stars with extreme colors
suitable for defining the color terms in the photometric transformation, the
calibration should be extremely well-defined for stars with the neutral colors
of Cepheid variables ($B-I \approx 1.3$--1.8 mag).

    The ACS measurements of these same 250 stars implied the following
transformation equations:

$$b_{\hbox{ACS}} = B + {\hbox{\it constant}} - 0.091 (\hbox{\it B--V\/}) + 0.075 (\hbox{\it B--V\/})^2,$$
$$v_{\hbox{ACS}} = V + {\hbox{\it constant}} + 0.018 (\hbox{\it V--I\/}) + 0.038 (\hbox{\it V--I\/})^2,~{\rm and}$$
$$i_{\hbox{ACS}} = I + {\hbox{\it constant}} + 0.088 (\hbox{\it V--I\/}) - 0.030 (\hbox{\it V--I\/})^2.$$

\noindent
Here we do not report the quantitative values of the zero-point constant
because they would not apply to ACS images drizzled differently from the way we
have done; the 47~Tuc images we employed had been drizzled in exactly the same
way as the NGC$\,$3370 images, and we transfered these zero-points as well as
the color transformations. The standard deviation of the 250 individual stars
about these mean relations was 0.015 mag in $B$ and $V$, and 0.026 mag in
$I$, so the contribution of random calibration uncertainty to the total error
budget should be very small (formally $< 0.002$ mag).  Accordingly, our
external errors should be dominated by any time variation in the ACS
throughput, and other systematics stemming from differences between the 47~Tuc
images and the NGC$\,$3370 images.

   This photometric solution was applied to 51 bright, isolated stars (Table 3)
intended to serve as local reference standards in the NGC$\,$3370 field.  This
list may be used in the future to test the zero-points or color dependencies of
our photometry.  The photometric solution was also applied to the faint stars
in NGC$\,$3370 yielding estimated stellar magnitudes in the Landolt (1992) version
of the Johnson/Cousins photometric system.  This is the same general procedure
used by the SKP (e.g., Hill et al. 1998), in that the SKP also derived
transformations of $HST$ data to Landolt's ground-based $V_J,I_C$ system from a
preliminary version of Stetson's homogeneous standard system.  The present
calibration is independent of that used by the SKP, in that theirs was based on
the clusters Pal~4 and NGC$\,$2419 and was employed, of course, for calibrating
WFPC2 rather than ACS.

   An independent route was to derive magnitudes for the field stars and those
in NGC$\,$3370 by utilizing the ``official'' photometric calibration obtained
by the ACS Team and STScI (Sirianni et al. 2005).  This calibration is based on
an empirical measurement of the system throughput at all wavelengths using the
known spectral-energy distributions of five white dwarfs.  A transformation
from the ACS natural system to the Johnson/Cousins system was derived by
producing synthetic magnitudes from spectrophotometry of a wide range of
stellar types.  This method is similar to the WFPC2 calibration by Holtzman et
al. (1995).  We will use the 51 local standards in the NGC$\,$3370 field to
compare the two methods of calibrations in \S 2.3.

   Comparisons of photometry from the same WFPC2 images for 118 Cepheids using
two popular PSF-fitting tools, DoPHOT and DAOPHOT, show that these methods
yield magnitudes which typically agree to within a few 0.01 mag (Gibson et
al. 2000; Saha et al.  2001).  Leonard et al. (2003) compared photometry
obtained by DAOPHOT/ALLFRAME and HSTphot [a version of DoPhot customized by
Dolphin (2000) for WFPC2] for nonvariable stars in NGC$\,$1637 and derived a
mean difference of $\Delta V = 0.015 \pm 0.022$ mag and $\Delta I = 0.036 \pm
0.018$ mag; for our purposes such differences are negligible.  Indeed,
according to Saha et al. (2001), the difference in distances measured from the
same data by SKP and STS lies not in the method used to fit the PSF but rather
in differences in the samples of Cepheids selected and how they are treated; we
thus concluded that it was not necessary to repeat measurements with DoPhot. We 
will explore the sensitivity to the analysis parameters in \S 2.2 and 2.3.

\subsection{Identification of Cepheid Variables}

    We used a number of selection criteria to identify Cepheids.  As in
Leonard et al. (2003), our criteria were governed by the philosophy that it is
better to fail to identify a real Cepheid than to include a non-Cepheid in the
sample.
  
   We required that all candidates have reported photometry for all twelve
epochs in $F555W$ and all five epochs in $F814W,$ to reduce the risk of aliases
in inadequately sampled light curves.  We required a modified Welch/Stetson
variability index\footnote{This is a ratio of apparent variation over expected
precision allowing some degree of correlation among magnitudes measured close
together in time (Welch \& Stetson 1993).  In the present instance, $F555W$ and
$F814$ magnitudes obtained during the same visits were paired, while the
remaining $F555W$ epochs were treated as individual, unmatched observations.}
in excess of 2.50, a threshold that corresponds to the 99.8 percentile among
the 56,244 measured sources that have all 17 mandatory magnitude measurements.
These criteria selected an initial sample of 96 Cepheid candidates from which
we rejected one, because it had $\left<V\right> > 28\,$mag, where the
photometric errors were judged to be excessively large for our purposes.  We
verified by inspection that the residuals from the PSF fits of these 95
detections were all consistent with an unresolved source.
   
   Stars meeting these criteria were subjected to a ``minimum string-length''
analysis.  This process identified as most likely the trial periods which
minimized the sum of magnitude variations for observations at similar phases
(Stetson et al. 1998).  In this way 20 of the most plausible periods were
determined for each candidate.  Robust least-squares fits were then performed,
comparing the single-epoch magnitudes to template Cepheid light curves, where
five parameters representing (1)~period, (2)~amplitude, (3)~epoch of zero
phase, (4)~mean magnitude in $V$, and (5)~mean magnitude in $I$ were free
parameters, but the shape of the light curve was a unique function of the
assumed period.  Templates were fit in $V$ and $I$ simultaneously, with a
single value of the period, phase, and amplitude required for both bandpasses,
but independent mean magnitudes for the two bands.  When the optimum fit had
been achieved, the fitted light curves were converted to flux units and
numerically integrated over a cycle to achieve flux-weighted mean luminosities
in each bandpass; these were then converted back to magnitude units.

   Finally, one of us (P.B.S.) applied his experienced judgment to a visual
comparison of the data to the adopted template to categorize the individual
light-curve fits as ``very good,'' ``good,'' ``fair,'' and ``poor.''
Judgment criteria included the following. (1) When the observed $V$ magnitudes
were plotted against a linear increase in epoch, they differed from each other
by amounts large compared to their individual measuring uncertainties, and a
consistent rise-and-fall pattern of one or more cycles was seen. (2) When
phased with the adopted period, the $V$ data were well distributed in phase and
showed the classic rapid-rise, slow-decline sawtooth light curve of
fundamental-mode radially pulsating variables. (3) All, or nearly all, of the
observed $V$ magnitudes agreed with the template light curve by amounts that
were consistent with their individual uncertainties. (4) To the extent that the
phase sampling allowed us to judge, the $I$-band magnitudes phased to an
appropriate light curve with epochs of maximum and minimum brightness
coinciding with those in $V$.

   This partially subjective judgment was checked by a purely numerical
process.  Following Leonard et al. (2003), from our empirically derived list of
Cepheids we selected those objects whose best-fitting template had a ``relative
likelihood'' greater than 0.3, where the ``relative likelihood'' is defined as
the quantity exp($-\chi^2/2\nu$), with $\nu$ representing the number of degrees
of freedom of the fit (in our case, 12 $V$-band epochs and 5 $I$-band epochs minus 
5 parameters in the fit yields $\nu = 12$, so this limit implies $\chi^2 < 30$).
Here, $\chi^2$ is the usual sum over all data points of the square of the model
residual normalized by the individual measurement uncertainty.  We also
implemented a more ``robust'' version (i.e., more forgiving) of the $\chi^2$
threshold by rejecting the single point (among the 17) making the largest
contribution to the $\chi^2$ sum to reflect the potential for imperfect
cosmic-ray rejection for 4 independent images per epoch.

   In all, 35 very ``good,'' 29 ``good,'' 13 ``fair,'' and 18 ``poor'' Cepheids
were identified by the visual inspection (by P.B.S.).  The aforementioned 95
variable candidates are thus reduced to 64 as either ``very good'' and ``good''
objects selected by inspection.  The method of selection by the $\chi^2$
criterion was more restrictive, identifying 31 and 51 Cepheids by the $\chi^2$
and robust $\chi^2$ criteria, respectively.  The 70 objects identified as
Cepheids by at least one of the methods are presented in Table 4 along with
their positions, photometric properties, and the criteria by which they were
selected.  Their $V$-band and $I$-band light curves are shown in Figure 5.  For
these same objects, their position in the color-magnitude ($V-I$ vs. $V$)
diagram of all detected unresolved sources is shown in Figure 6.  As expected,
the objects identified as Cepheids have $V-I \approx 1.0$ mag, consistent with
stars crossing the instability strip.

   Obtaining accurate photometry of Cepheids on the crowded and granular
background of their hosts is challenging.  However, the impact of crowding on
the measured magnitudes has been largely eliminated by application of the
previously described selection criteria, whose effect has been to disfavor
Cepheids that are significantly contaminated (e.g., by a close binary
companion).  Monte Carlo experiments by the Key Project (Ferrarese et al. 2000)
have shown (by reproducing the effects of crowding with artificial PSFs) that
the presence of significant contamination will alter the shape parameters of
the Cepheid PSF, reducing the amplitude of variation and flattening the
sawtooth near minimum light (by contributing a greater fraction to the total
flux when the Cepheid is faint).  These changes will typically cause a Cepheid
candidate to fail one or more of the previous criteria.  Ferrarese et
al. (2000) found that for multi-epoch data, the net crowding bias on the
distance modulus is only $\sim$1\%.  In practice, we found that nearly all of
the candidates in locations judged to be ``crowded'' (defined here as having an
additional source within at least $0.1''$ which contributes at least $\sim$10\%
of the peak flux of the variable candidate) failed one or more of the
previously discussed selection criteria.

\clearpage

\begin{deluxetable}{lllll} 
\footnotesize 
\tablenum{2}
\tablecaption{$HST$ Observations of NGC 3370} 
\tablehead{\colhead{Epoch}&\colhead{$HST$ Archive rootname}&\colhead{UT Date}&\colhead{MJD$^a$ at Start}&\colhead{Exp. Time (s)}} 
\startdata \multicolumn{5}{c}{$F555W$} \nl 
\hline 1 & j8d211dzq  &      2003-03-31   &    52729.51097335  &     4800.0 \nl
 2 &  j8d212pdq  &      2003-04-08   &    52737.58257076  &     4800 \nl
 3 &  j8d213giq  &      2003-04-16  &     52745.38616072  &     4800 \nl
 4 &  j8d214f5q  &      2003-04-21  &     52750.92149627  &     4800 \nl
 5 &  j8d215neq  &      2003-04-27  &     52756.12557049  &     4800 \nl
 6 &  j8d216p4q  &      2003-04-29  &     52758.05614901  &     4800 \nl
 7 &  j8d217a6q  &      2003-05-01  &     52760.12780428  &     4800 \nl
 8 &  j8d218m8q  &      2003-05-03  &     52762.05838317  &     4800 \nl
 9 &  j8d219v7q   &     2003-05-07   &    52766.13153132  &     4800 \nl
 10 &  j8d21amgq   &     2003-05-09 &     52768.41565169  &     4800 \nl
 11 &  j8d21bbqq  &      2003-05-15  &    52774.40382280  &     4800 \nl
 12 &  j8d21capq   &     2003-05-22 &     52781.41830234  &     4800 \nl
\hline
\multicolumn{5}{c}{$F814W$} \nl
\hline
 1 &  j8d211etq   &     2003-03-31  &     52729.64275354   &    4800 \nl
 3 &  j8d213gzq   &     2003-04-16  &     52745.51695924  &     4800 \nl
 7 &  j8d217avq   &     2003-05-01  &     52760.33845243  &     4800 \nl
 10 &  j8d21amyq   &     2003-05-09  &    52768.54504984  &     4800 \nl
 12 &  j8d21cauq   &     2003-05-22  &    52781.46974901  &     4800 \nl
   \enddata 
\tablenotetext{a}{MJD is the Julian date minus 2400000.}
\end{deluxetable}

\clearpage

\begin{deluxetable}{ccccc}
\footnotesize
\tablenum{3}
\tablecaption{Comparison Stars in ACS Field of NGC 3370, truncated for astro-ph}
\tablehead{\colhead{ID}&\colhead{$x^a$}&\colhead{$y^a$}&\colhead{$<V>,\sigma^b$}&\colhead{$<I>,\sigma^b$}}'
\startdata
 349.0 & 326.25 & 202.78 & 22.935  0.006 & 22.137  0.010 \nl
\enddata
\tablenotetext{a}{$x$ and $y$ positions are for pixels of $0.033''$ per pixel
as observed in the epoch 1 image with the origin at $x=153$, $y=137$
in the detector orientation.}
\tablenotetext{b}{$V$-band and $I$-band magnitudes have been transformed to the
Johnson/Cousins system. Uncertainties are $1\sigma$.}
\end{deluxetable}

\clearpage

\begin{deluxetable}{ccccccc}
\footnotesize
\tablenum{4}
\tablecaption{Cepheid Candidates, truncated for astro-ph}
\tablehead{\colhead{ID}&\colhead{$x^a$}&\colhead{$y^a$}&\colhead{P (d)}&\colhead{$<V>,\sigma^b$}
&\colhead{$<I>,\sigma^b$}&\colhead{$\chi^2 / robust / PBS$ }}
\startdata
 71916 & 3109.0 & 3585.7 & 54.0 & 26.54  0.02 & 25.47  0.01 & NNY\nl
\enddata
\tablenotetext{a}{$x$ and $y$ positions are for pixels of $0.033''$ per pixel
as observed in the epoch 1 image with the origin at $x=153$, $y=137$ 
in the detector orientation.}
\tablenotetext{b}{$V$-band and $I$-band magnitudes have been transformed to the
Johnson/Cousins system. Uncertainties are $1\sigma$.}
\end{deluxetable}

\clearpage

\subsection{The Cepheid Distance to NGC 3370}

   In principle, the average luminosity of a Cepheid variable is accurately
identified from its period, $P$, via the simple relation
$$ M_c=a_c \log (P)+b_c,$$ 
where $M_c$ is the intensity-mean absolute magnitude of a Cepheid in a passband
$C$, and $a_c$ and $b_c$ are the slope and zero-point of the relation.  The
apparent distance modulus is then given by the apparent and absolute magnitudes,
$$\mu_C = m_C - M_c.$$  
A modest correction to this simple picture is required to allow for the nonzero
width of the instability strip: Cepheids of a given luminosity do not
necessarily have identically the same radii or, hence, the same effective
temperatures.  This produces a period-luminosity-color relation among classical
Cepheids.  The need for a final small correction to account for differences in
chemical abundance is addressed in \S 2.5.

   The use of two or more different passbands allows for the measurement of
reddening and the associated correction for extinction to give the true or
unreddened distance modulus [also called the ``Wesenheit reddening-free
modulus'' (Madore 1982; Tanvir 1997)].  For passbands $V$ and $I$ this is given
as $\mu_W = \mu_V-R(\mu_V-\mu_I)$, where $R = A_V / (A_V-A_I)$ is typically
taken to be 2.45 as derived from the wavelength-dependent extinction curve of
Cardelli, Clayton, \& Mathis (1989).  It is fortunate that this is almost
exactly the same slope as the relationship between color and magnitude for
Cepheids (i.e., for dilute blackbodies) of differing effective temperature at
fixed period.  Therefore, the reddening correction simultaneously and nearly
perfectly removes the complication produced by the nonzero width of the
instability strip.

   However, the quantitative determination of the best values for the slope and
zero-point of the resulting period-luminosity (hereafter \PL; no color) relation
is still subject to ongoing refinement and remains a subject of debate,
contributing to past differences between the SKP and the STS groups.  The
determination of its zero-points also suffers from the uncertainty in the
distance and reddening to the LMC, which hosts the
largest observed sample of Cepheids.  Here we have utilized a small set of the
most well-founded and likely values of the \PL relations.

   Most of the Cepheid analyses by the SKP and the STS groups in the 1990s was
referenced to the sample of 32 Cepheids with photoelectric data in the LMC and
$1.6 < P < 63$ days as compiled by Madore \& Freedman (1991; hereafter MF91).
However, a significant improvement in the available sample of LMC Cepheids was
later made by the Optical Gravitational Lensing Experiment (OGLE; Udalski et
al.  1999), which includes $\sim$650 Cepheids with periods in the range $2.5 <
P < 31$ days and an extremely high rate of data sampling. The photometry from
these measurements has been independently verified by Sebo et al. (2002), who
also extended the sample to include periods up to 40 days.  In their final
analysis the SKP adopted the \PL relations derived from the OGLE data.
However, a valid criticism of this sample is that $\sim$90\% of the OGLE
Cepheids have $P < 10$ days, which is significantly shorter than for the
faintest Cepheids visible with $HST$ in the hosts of SNe~Ia.  A more
appropriate \PL relation was derived by Thim et al. (2003) and Tammann \&
Reindl (2002) by limiting the OGLE Cepheids to the 44 with $P > 10$ days.  To
provide a zero-point of these relations for our own analysis, we adopt the
canonical distance to the LMC of $\mu = 18.50 \pm 0.1$ mag as currently used by
the SKP group (F01) and the STS group (Saha et al. 2001;
however, Tammann et al. 2001 more recently prefer 18.56 mag and the LMC distance modulus ).
  Here we consider the
\PL relation from only the longer-period OGLE data chosen by Thim et
al. (2003) as our preferred reference sample because of its virtues of
utilizing the high-quality, photometrically tested OGLE data while providing a
better match to the long-period Cepheids identified with $HST$.  All three
aforementioned \PL relations are given in Table 5.

   In Figure 7 we show the $V$-band and $I$-band \PL relations of the
Cepheids in NGC 3370.

\clearpage

\begin{deluxetable}{lrrrrrrl}
\small
\tablenum{5}
\tablewidth{0pt}
\tablecaption{Slopes and Zero-Points for Cepheid \PL Relations}
\tablehead{\colhead{Set} &
\colhead{$a_V$} &
\colhead{$a_I$} &
\colhead{$a_W$} &
\colhead{$b_V$}  &
\colhead{$b_I$} &
\colhead{$b_W$} &
\colhead{Reference}}

\startdata
OGLE & $-2.760$ & $-2.962$ & $-3.255$ & $-1.458$ & $-1.942$ & $ -2.644$ & 1\\
MF91 & $-2.760$ & $-3.060$ & $-3.495$ & $-1.400$ & $-1.810$ & $ -2.405$ & 2\\
OGLE+10 & $-2.480$ & $-2.820$ & $-3.313$ & $-1.750$ & $-2.090$ & $ -2.583$ & 3\\
\enddata

\tablecomments{Slopes and zero-points for the \PL relations resulting from the
four samples of Cepheids described in \S 2.3.  The relations are defined by
$M_C = a_C \log (P) + b_C$, with $M_C$ the absolute magnitude of a Cepheid in
photometric band $C$, $P$ its period (days), and $a_C$ and $b_C$ the respective
slope and zero-point of the relation. $W$ represents the Wesenheit
reddening-free index described in the text.}

\tablerefs{ (1) Freedman et al. 2001; (2) Madore \& Freedman 1991; (3) Thim et al. 2003.}

\label{tab:tab1}

\end{deluxetable}

\clearpage

\begin{deluxetable}{cccc} 
\footnotesize
\tablenum{6}
\tablewidth{0pt}
\tablecaption{Cepheid-Based Distances to NGC 3370}
\tablehead{\colhead{Zero-point}&\colhead{       }&\colhead{\PL Relations}&\colhead{    }}
\startdata
\multicolumn{1}{c}{ } & \multicolumn{1}{c}{MF91} & \multicolumn{1}{c}{OGLE}& \multicolumn{1}{c}{OGLE+10} \nl 
\hline
\multicolumn{1}{c}{PBS selection} \nl
Stetson & 32.28(0.03) & 32.13(0.03) & 32.17(0.03) \nl
Sirianni & 32.33(0.03) & 32.19(0.03) & 32.23(0.03) \nl
\multicolumn{1}{c}{$\chi^2$ selection} \nl
Stetson & 32.32(0.03) & 32.18(0.03) & 32.21(0.03) \nl
Sirianni & 32.38(0.03) & 32.24(0.03) & 32.27(0.03) \nl
\multicolumn{1}{c}{robust $\chi^2$ selection} \nl
Stetson & 32.32(0.04) & 32.17(0.04) & 32.20(0.04) \nl
Sirianni & 32.37(0.04) & 32.23(0.04) & 32.26(0.04) \nl
\hline
\enddata
\end{deluxetable}

\clearpage

   In principle, the mean magnitudes of shorter-period (e.g., $P < 20$ days),
fainter Cepheids ($F555W > 27.0$ mag) can be biased by their selection from a
magnitude-limited sample.  However, the shortest-period Cepheid selected by any
of our selection methods, $P = 17.5$ days, is still detected at S/N $\approx
14$ at each $F555W$ epoch.  Such a Cepheid whose apparent magnitude was up to
2.0 standard deviations fainter than average by chance (which includes 97.5\%
of the sample) could still be detected at S/N $\approx 9$ at each epoch.
Consequently, no selection bias would be expected for our sample.  As an
additional test, we measured the distance to NGC$\,$3370 as a function of a
minimum period cutoff.  As seen in Figure 8, the modulus does not significantly
increase for increasing the minimum cutoff as would be expected under the
influence of a magnitude-limit bias.  Therefore, we impose no minimum period
cutoff on our sample (higher than the lowest detected period, 17.5 days).

   In Table 6 we have utilized the aforementioned Cepheid identification
methods, the two independent photometric zero-point determinations, and the
three alternative \PL relations to determine the distance to NGC$\,$3370.  The
combination which most closely matches the procedures of the SKP is the use
of the visual classification, the Stetson zero-points, and the \PL relation from
Freedman et al. (2001; hereafter F01).  This combination results in a distance
modulus of $\mu_w = 32.13 \pm 0.03$ mag.  Alternatively, we can reproduce the
distance modulus which would be achieved by the STS collaboration by selecting
the Sirianni zero-points.   The Sirianni et al. (2005) zeropoints for ACS are a good proxy for the STS zeropoints because they are similarly derived empirically from the HST throughput and in addition Sirianni et al. (2005) have shown that their ACS zeropoints are consistent with the Saha et al. (2001) WFPC2
zeropoints (for similar fields) to a precision of 0.01 mag.  To emulate
STS we also use the  pre-OGLE \PL relations from MF91.  These
selections yield a range of $\mu_w = 32.33$--32.38 mag for the three Cepheid
selection criteria; hereafter we will assume an average of $\mu_w = 32.36$ mag.

   However, our goal is not simply to repeat the methodology of these teams but
rather to choose what we consider to be the optimal combination of parameters
based on the most current available information.  It is apparent from the
sample size and time-sampling of the OGLE Cepheids, as well as the verification
by Sebo et al.  (2002) of their photometric accuracy, that these data are now
preferred in defining the \PL relation. Moreover, in order to compare Cepheids
with similar periods free from concerns of a discontinuity in the \PL relation
expressed by Tammann \& Reidel (2002), we use the Thim et al. (2003) \PL
relation, which utilizes the OGLE data for $P > 10$ days.  From this we get
$\mu_w = 32.17 \pm 0.03$ to $32.27 \pm 0.03$ mag (where the uncertainties
listed are statistical).  We will use $\mu_w = 32.22 \pm 0.05$ mag as our best
estimate, which is representative of this range but includes a greater
uncertainty to account for the systematic uncertainty resulting from the
different analyses.

\subsection{Cepheid Distances to Other SN Ia Hosts}

   In order to provide a consistent Cepheid-based distance to NGC$\,$3982, the
host of SN 1998aq (whose photometry is presented here for the first time in \S
3), we have revisited the analysis of the Cepheids by both the STS and SKP
teams (Saha et al. 2001; Stetson \& Gibson 2001).  The analyses of both teams
are careful and thorough, and both provide similar tables of discovered
Cepheids and their photometric parameters.  Saha et al. conclude that $\mu_0 =
\mu_w = 31.72 \pm 0.14$ mag, while Stetson \& Gibson conclude that $\mu_0 =
\mu_w = 31.56 \pm 0.08$ mag (considering only statistical uncertainties).  As
noted by Stetson \& Gibson and confirmed here, despite the independence of the
analysis, the primary cause for the difference in distance modulus is from the
choice of \PL relations; Stetson \& Gibson use the same OGLE-based relation as
F01, while Saha et al. use the earlier relation from MF91.  Repeating the
analysis of the Cepheids observed in NGC$\,$3982 as reported by the two groups,
but now using the \PL relation from the OGLE+10 Cepheids (Thim et al. 2003), we
find $\mu_w = 31.60 \pm 0.08$ mag.

   As previously discussed, there are two additional SNe~Ia (SN 1990N and SN
1981B) whose characteristics and data quality match the modern sample of
high-quality objects defining the Hubble flow and thus can provide calibrations
of the fiducial luminosity of SNe~Ia as good as those from SN 1998aq and SN
1994ae.  Cepheids in their hosts, NGC 4639 (SN 1990N) and NGC 4536 (SN 1981B),
were observed by the STS collaboration and were analyzed by both STS (Saha et
al. 1996, 1997) and SKP (Gibson et al. 2000).  For consistency we have refit
the published photometry of these Cepheids to relations based on the OGLE+10
Cepheids as above. We used the photometry of Gibson et al. (2000) for NGC 4536
which, unlike that of Saha et al. (1996), finds consistency between the moduli
from WFPC2 Chip 2 and that from the other 3 chips.  For NGC 4536 we find $\mu_0 = 30.78
\pm 0.07$ mag and for NGC 4639 we find $\mu_0 = 31.62 \pm 0.09$ mag.

\subsection{Metallicity}

   Mounting evidence indicates that there is a significant dependence of the
apparent magnitudes of Cepheids (at fixed period) on the metallicity of the
host galaxy.  A marginal (1.5$\sigma$) detection of this dependence by
Kennicutt et al. (1998) was derived from two different Cepheid fields in M101
differing in [O/H] by 0.7 dex.  A greatly improved calibration of the
dependence (consistent with that used by the Key Project) was derived by Sakai
et al. (2004) using 17 Cepheid hosts with independent distances from the
apparent magnitudes of the tip of the red giant branch.  This calibration, 
$\delta (m-M) / \delta{\rm [O/H]} = -0.24 \pm 0.05$ mag dex$^{-1}$, is used here to
correct the SN~Ia host Cepheids for their metallicity difference with the LMC.

   The metallicities of the two SN hosts, NGC$\,$3370 and NGC$\,$3982, are not
available in the literature.  To obtain estimates we followed the procedure
established by Zaritsky, Kennicutt, \& Huchra (1994) and subsequently utilized
by Kennicutt et al. (1998) and Sakai et al. (2004).  Slit masks were produced
for the Low Resolution Imaging Spectrometer (Oke et al. 1995) on the
Keck-I telescope for likely H~II regions based on the archival $HST$ and
ground-based (Lick Observatory 1-m ``Nickel telescope'') images of the two 
galaxies observed through narrow-band H$\alpha$ and broad-band
filters.  Emission-line intensity ratios of [O~II] $\lambda\lambda$3726, 3729,
[O~III] $\lambda\lambda$4959, 5007, and H$\beta$ were measured from our
resulting spectroscopy of the H~II regions.  The line ratios were corrected for
reddening as derived from the Balmer decrement (assuming the unreddened values
from Case B recombination; Osterbrock 1989).  For each H~II region, the
$R_{23}$ value [defined as ([O~II] + [O~III])/H$\beta$] and values for 12 +
log(O/H) were derived from the empirical calibration from Zaritsky et
al. (1994) which assumes [O/H]$_{solar} = 7.9 \times 10^{-4}$.  For NGC 3370 and NGC
3982, 10 and 7 independent H~II regions, respectively, were measured and their
metallicities calibrated.

   A strong, linear gradient in metallicity with distance from the nucleus of
each host is seen, as expected, with a dispersion of $\sim$0.1 dex.  The mean
metallicity at the same average galactocentric radius as the $HST$-observed
Cepheids of $50''$ for both galaxies was inferred from a linear fit to the
gradients shown in Figure 9, the results of which are given in Table 7.

\clearpage

\begin{deluxetable}{cccc} 
\footnotesize
\tablenum{7}
\tablecaption{Metallicities of SN~Ia Hosts}
\tablehead{\colhead{Host}&\colhead{12 + log(O/H)}&\colhead{[O/H]--[O/H]$_{\rm
LMC}$} & \colhead{$\delta(m-M)$ (mag)}}
\startdata
NGC 3370 &  $8.80 \pm 0.05$ & $0.30 \pm 0.05 $  &$0.07 \pm 0.03$  \nl
NGC 3982 &  $8.75 \pm 0.05$ & $0.25 \pm 0.05 $  &  $0.06 \pm 0.03$ \nl
NGC 4639$^a$ &  $9.00 \pm 0.20$ & $ 0.50 \pm 0.20$  &  $0.12 \pm 0.08$ \nl
NGC 4536$^a$ &  $8.85 \pm 0.20$ & $ 0.35 \pm 0.20$  &  $0.08 \pm 0.05$ \nl
\enddata
\tablenotetext{a}{From Gibson et al. 2000.}
\end{deluxetable}

\clearpage

\section{The Light Curves of SN 1994ae and SN 1998aq}

    SN 1994ae and SN 1998aq were both spectroscopically normal SNe~Ia.  SN
1994ae in NGC$\,$3370 was discovered (Van Dyk et al. 1994) on 3 Nov. 1994 (UT
dates are used throughout this paper) by the Leuschner Observatory Supernova
Search (Treffers et al. 1995) with the Berkeley Automatic Imaging Telescope
(Treffers, Richmond, \& Filippenko 1992; Richmond, Treffers, \& Filippenko
1993).  At that time, it was $\sim$12 days before maximum brightness, earlier
than nearly all previous SN~Ia discoveries.  Its location in the galaxy was
6\Sec1 north and 29\Sec7 west as seen in Figure 2.  Photometric monitoring of
the light curve commenced immediately by the CfA Survey I (Riess et al. 1999a).
An identification spectrum was obtained by Iijima, Cappellaro, \& Turatto
(1994), and additional spectroscopy was obtained at the Mt. Hopkins 1.5-m
telescope (Matheson et al. 2005), revealing typical SN~Ia features including
strong, broad, blueshifted Si~II and S~II absorption.

  SN 1998aq in NGC$\,$3982 was discovered on 13 April 1998 by M. Armstrong
during the course of the U.K. Nova/Supernova Patrol, $18''$ west and $7''$
north of the center of NGC 3982 (Hurst, Armstrong, \& Arbour 1998).  The
original CCD discovery images have been transformed to standard passband
systems by Riess et al. (1999b).  Photometric monitoring of the SN commenced
immediately at the FLWO 1.2-m telescope by the CfA Survey II (Jha et
al. 2005a).  An identification spectrum was obtained on 15 April by Ayani \&
Yamaoka (1998), revealing broad, blueshifted absorption of Si~II and S~II
consistent with a typical SN~Ia.  A comprehensive spectroscopic record of SN
1998aq was published by Branch et al. (2003).
    
   The relative proximity of both of these SNe~Ia ($d < 30$ Mpc and $V_{\rm
max} \leq 13.0$ mag) provides two rare and valuable opportunities to improve
the calibration of the fiducial luminosity of SNe~Ia using the techniques and
tools of modern photometry.  Both SNe are ideal calibrators by the criteria
defined in \S 1.
   
   Initial photometry was published for SN 1994ae by Riess et
al. (1999a), but this reduction lacked the advantages of galaxy
template subtraction and repeated zero-point calibrations.  Here we
have undertaken an improved calibration of SN 1994ae by performing
galaxy subtraction and repeating the calibration of the field stars
four times. For SN 1998aq, we present the photometry here for the
first time.

   For SN 1994ae, photometric monitoring was conducted with the FLWO 1.2-m
telescope, an $f/8$ Ritchey-Chr\'{e}tien reflector equipped with a thick
front-illuminated Loral CCD and a set of Johnson $UBV$ and Kron-Cousins $RI$
passbands (hereafter the ``SAO'' filter set).  Further details of the
equipment, typical color terms, and broad-band transmission functions are
provided by Riess et al. (1999a), and updates to the equipment at this facility
are described by Jha et al. (2005a).  On nights which were judged to be
photometric, we performed all-sky photometry and derived a transformation from
our detector measurements to the standard photometric conventions using Landolt
(1992) standards stars.  From each useful transformation we derived the
magnitudes of comparison stars in the field of the supernova.  For SN 1994ae, 3
photometric transformations (7 Jan. 1995, 8 Dec. 1999, and 15 Feb.  2000) were
derived from the 1.2-m telescope, and two more (31 May 2003 and 1 June 2003)
came from the Lick Observatory 1-m Nickel telescope (see Ho et al. 2001 for
average color terms).  Average magnitudes for field stars are given in Table 8
and a finder chart for these stars is shown in Figure 10.  Galaxy templates
were selected from the FLWO imaging on 8 Dec. 1999.

   In general, the same procedures as described for SN 1994ae were followed for
SN 1998aq.  Photometric monitoring was conducted with the FLWO 1.2-m telescope
using the same SAO filter set as for SN 1994ae.  During this monitoring, one of
two thinned, backside illuminated, anti-reflection coated Loral CCD detectors
was used, primarily the ``4Shooter'' and on two dates the ``AndyCam'' (MJD
$- 2450000$ = 960 and 1013); for more detailed specifications for these cameras
see Jha et al. (2005a). Average magnitudes for field stars are given in Table 9
and a finder chart for these stars is shown in Figure 14.  For SN 1998aq, five
independent transformations were used from the nights of 31 May 2003 (Nickel),
15 and 22 June 1998 (FLWO), and 19 and 26 March 2001 (FLWO).  Galaxy
templates were obtained from imaging at the FLWO on 27 Nov. 2000.

  Supernova magnitudes were determined from the relative photometry of the
comparison stars and the SN after galaxy subtraction.  The PSF-fitting
procedure in IRAF was used to conduct the photometry.  Model PSFs were
constructed for each image using the bright, isolated stars in the frame and
were then used to fit the magnitudes of the SN and the comparison stars.
Average color terms (given in Riess et al. 1999a and Jha et al. 2005a) were used 
to account for color differences between the SN and comparison stars, and to
transform the SN magnitudes to the Johnson $UBV$ and Kron-Cousins $RI$
systems.\footnote{A more detailed accounting for the effect of the nonstellar
spectral energy distribution of SNe~Ia on transformed magnitudes (i.e.,
``S-corrections''; Suntzeff et al. 1999) is easily accomplished by returning
the magnitudes to the natural system (using the average color terms), and
making use of the transmission functions in Riess et al. (1999a) and Jha et
al. (2005a) in addition to template SN~Ia spectra to transform the data to
other passband conventions.}  The final photometry for SN 1994ae and SN 1998aq
is listed in Tables 10 and 11, respectively.  The uncertainty of the
measurements was determined as the quadrature sum of the individual PSF fit
uncertainty and the standard deviation of the transformation from the
comparison stars.

\clearpage

\begin{deluxetable}{cccccc}
\footnotesize
\tablenum{8}
\tablecaption{Comparison Stars for SN 1994ae}
\tablehead{\colhead{Star}&\colhead{$U$($\sigma$)N}&\colhead{$B$($\sigma$)N}&\colhead{$V$($\sigma$)N}&\colhead{$R$($\sigma$)N}&\colhead{$I$($\sigma$)N}}
\startdata

01 & 18.214 ... 1 & 18.402 0.022 4 & 17.811 0.033 3 & 17.443 0.045 2 & 17.027 0.064 3 \nl
\enddata
\end{deluxetable}

\clearpage

\begin{deluxetable}{cccccc}
\footnotesize
\tablenum{9}
\tablecaption{Comparison Stars for SN 1998aq}
\tablehead{\colhead{Star}&\colhead{$U$($\sigma$)N}&\colhead{$B$($\sigma$)N}&\colhead{$V$($\sigma$)N}&\colhead{$R$($\sigma$)N}&\colhead{$I$($\sigma$)N}}
\startdata
   
   1  & 17.337  0.005   3  & 16.180  0.027   4  & 14.926  0.025   4  & 14.126  0.017   4  & 13.424  0.029   4 \nl
\enddata
\end{deluxetable}

\clearpage

\begin{deluxetable}{cccccc}
\footnotesize
\tablenum{10}
\tablecaption{Photometry of SN 1994ae}
\tablehead{\colhead{MJD$^a$}&\colhead{$U$($\sigma$)}&\colhead{$B$($\sigma$)}&\colhead{$V$($\sigma$)}&\colhead{$R$($\sigma$)}&\colhead{$I$($\sigma$)}}
\startdata
      9672.97& ... & 14.865 ( 0.041 ) & 14.833 ( 0.027 ) &
 14.711 ( 0.022 ) & 14.599 ( 0.028 ) \nl
      9674.96& ... & 14.109 ( 0.044 ) & 14.133 ( 0.040 ) &
 14.007 ( 0.026 ) & 13.948 ( 0.019 ) \nl
      9676.96& ... & 13.730 ( 0.167 ) & 13.690 ( 0.051 ) &
 13.580 ( 0.019 ) & 13.590 ( 0.033 ) \nl
      9678.00& ... & 13.516 ( 0.030 ) & 13.513 ( 0.039 ) &
 13.422 ( 0.018 ) & ... \nl
      9686.00& 12.548 ( 0.084 ) & 13.059 ( 0.024 ) & 13.045 ( 0.030 ) &
 13.046 ( 0.028 ) & 13.287 ( 0.039 ) \nl
      9686.99& 12.608 ( 0.044 ) & 13.082 ( 0.028 ) & 13.077 ( 0.021 ) &
 13.064 ( 0.018 ) & 13.322 ( 0.031 ) \nl
      9688.04& ... & 13.104 ( 0.035 ) & 13.118 ( 0.030 ) &
 13.097 ( 0.015 ) & 13.363 ( 0.042 ) \nl
      9689.06& 12.728 ( 0.052 ) & 13.190 ( 0.047 ) & 13.120 ( 0.024 ) &
 13.097 ( 0.100 ) & 13.383 ( 0.070 ) \nl
      9690.03& 12.753 ( 0.052 ) & 13.231 ( 0.040 ) & 13.141 ( 0.036 ) &
 13.133 ( 0.034 ) & 13.449 ( 0.083 ) \nl
      9694.03& ... & 13.435 ( 0.038 ) & 13.306 ( 0.027 ) &
 13.312 ( 0.023 ) & 13.683 ( 0.026 ) \nl
      9695.03& ... & 13.515 ( 0.031 ) & 13.329 ( 0.034 ) &
 13.358 ( 0.020 ) & 13.784 ( 0.099 ) \nl
      9695.97& ... & 13.576 ( 0.035 ) & 13.398 ( 0.026 ) &
 13.438 ( 0.013 ) & 13.812 ( 0.022 ) \nl
      9696.96& ... & 13.674 ( 0.028 ) & 13.438 ( 0.028 ) &
 13.525 ( 0.015 ) & 13.877 ( 0.023 ) \nl
      9699.98& ... & 13.931 ( 0.046 ) & 13.645 ( 0.023 ) &
 13.715 ( 0.015 ) & 13.957 ( 0.031 ) \nl
      9700.97& ... & 14.048 ( 0.069 ) & 13.673 ( 0.024 ) &
 13.763 ( 0.023 ) & 14.001 ( 0.031 ) \nl
      9702.96& ... & 14.260 ( 0.034 ) & 13.785 ( 0.022 ) &
 13.793 ( 0.018 ) & 13.943 ( 0.030 ) \nl
      9705.06& ... & 14.446 ( 0.047 ) & 13.888 ( 0.028 ) &
 13.809 ( 0.020 ) & 13.890 ( 0.024 ) \nl
      9707.06& ... & 14.669 ( 0.040 ) & 13.960 ( 0.040 ) &
 13.836 ( 0.030 ) & 13.886 ( 0.060 ) \nl
      9715.93& ... & 15.491 ( 0.042 ) & 14.418 ( 0.029 ) &
 14.073 ( 0.040 ) & 13.773 ( 0.070 ) \nl
      9721.03& ... & 15.790 ( 0.040 ) & 14.677 ( 0.059 ) &
 14.382 ( 0.020 ) & 13.998 ( 0.036 ) \nl
      9725.01& ... & 15.949 ( 0.036 ) & 14.942 ( 0.022 ) &
 14.636 ( 0.019 ) & 14.256 ( 0.024 ) \nl
      9726.96& 16.113 ( 0.196 ) & 15.996 ( 0.060 ) & 14.967 ( 0.018 ) &
 14.685 ( 0.020 ) & 14.390 ( 0.027 ) \nl
      9740.86& ... & 16.262 ( 0.042 ) & 15.447 ( 0.030 ) &
 15.249 ( 0.035 ) & 15.048 ( 0.030 ) \nl
      9744.94& ... & 16.340 ( 0.041 ) & 15.541 ( 0.032 ) &
 15.342 ( 0.022 ) & 15.202 ( 0.032 ) \nl
      9745.92& ... & 16.348 ( 0.050 ) & 15.589 ( 0.026 ) &
 15.378 ( 0.023 ) & 15.263 ( 0.033 ) \nl
      9749.99& ... & 16.381 ( 0.042 ) & 15.650 ( 0.052 ) &
 ... & ... \nl
      9769.74& ... & 16.645 ( 0.070 ) & 16.191 ( 0.058 ) &
 16.136 ( 0.063 ) & 16.214 ( 0.070 ) \nl
      9783.86& ... & 16.861 ( 0.063 ) & 16.541 ( 0.069 ) &
 16.563 ( 0.072 ) & ... \nl
      9805.88& ... & 17.235 ( 0.082 ) & 16.982 ( 0.083 ) &
 17.186 ( 0.071 ) & 17.362 ( 0.080 ) \nl
\enddata
\tablenotetext{a}{MJD is the Julian date minus 2440000.}
\end{deluxetable}

\clearpage

\begin{deluxetable}{cccccc}
\footnotesize
\tablenum{11}
\tablecaption{Photometry of SN 1998aq}
\tablehead{\colhead{MJD$^a$} & \colhead{$U$($\sigma$)} &
\colhead{$B$($\sigma$)} & \colhead{$V$($\sigma$)} & \colhead{$R$($\sigma$)} &
\colhead{$I$($\sigma$)}}
\startdata
       920.64500& 12.724 ( 0.017 ) & 13.551 ( 0.027 ) & 13.708 ( 0.019 ) &
 13.613 ( 0.017 ) & 13.607 ( 0.015 ) \nl
       921.64300& 12.414 ( 0.016 ) & 13.264 ( 0.022 ) & 13.431 ( 0.016 ) &
 13.328 ( 0.016 ) & 13.359 ( 0.024 ) \nl
       922.63800& 12.202 ( 0.016 ) & 13.046 ( 0.022 ) & 13.208 ( 0.012 ) &
 13.110 ( 0.011 ) & 13.160 ( 0.019 ) \nl
       923.66100& 12.030 ( 0.020 ) & 12.861 ( 0.019 ) & 13.036 ( 0.013 ) &
 12.930 ( 0.013 ) & 13.010 ( 0.014 ) \nl
       923.66200& 12.031 ( 0.026 ) & 12.867 ( 0.020 ) & 13.029 ( 0.014 ) &
 12.929 ( 0.016 ) & 13.005 ( 0.021 ) \nl
       924.65000& 11.921 ( 0.015 ) & 12.729 ( 0.021 ) & 12.881 ( 0.018 ) &
 12.788 ( 0.014 ) & 12.879 ( 0.024 ) \nl
       924.65100& 11.916 ( 0.019 ) & 12.718 ( 0.030 ) & 12.885 ( 0.019 ) &
 12.790 ( 0.014 ) & 12.874 ( 0.026 ) \nl
       925.77400& 11.812 ( 0.014 ) & 12.600 ( 0.030 ) & 12.766 ( 0.025 ) &
 12.670 ( 0.026 ) & 12.776 ( 0.023 ) \nl
       925.77700& 11.816 ( 0.010 ) & 12.605 ( 0.019 ) & 12.777 ( 0.015 ) &
 12.661 ( 0.027 ) & 12.774 ( 0.026 ) \nl
       925.78000& ... & ... & 12.752 ( 0.023 ) &
 ... & ... \nl
       925.79000& ... & ... & 12.754 ( 0.028 ) &
 ... & ... \nl
       925.79100& ... & 12.596 ( 0.043 ) & 12.763 ( 0.030 ) &
 12.661 ( 0.021 ) & 12.775 ( 0.017 ) \nl
       926.74800& 11.732 ( 0.065 ) & 12.523 ( 0.047 ) & 12.666 ( 0.025 ) &
 12.602 ( 0.019 ) & 12.737 ( 0.022 ) \nl
       926.74800& 11.745 ( 0.049 ) & 12.522 ( 0.043 ) & 12.666 ( 0.020 ) &
 12.615 ( 0.051 ) & 12.745 ( 0.022 ) \nl
       928.75500& 11.715 ( 0.015 ) & 12.419 ( 0.027 ) & 12.531 ( 0.019 ) &
 12.501 ( 0.019 ) & 12.699 ( 0.024 ) \nl
       928.75500& 11.704 ( 0.032 ) & 12.410 ( 0.029 ) & 12.537 ( 0.022 ) &
 12.486 ( 0.025 ) & 12.701 ( 0.019 ) \nl
       930.77000& 11.723 ( 0.026 ) & 12.357 ( 0.028 ) & 12.479 ( 0.021 ) &
 12.445 ( 0.014 ) & 12.747 ( 0.021 ) \nl
       930.77100& 11.736 ( 0.043 ) & 12.358 ( 0.028 ) & 12.469 ( 0.015 ) &
 12.446 ( 0.014 ) & 12.743 ( 0.021 ) \nl
       931.79700& 11.785 ( 0.045 ) & 12.365 ( 0.017 ) & 12.458 ( 0.019 ) &
 12.434 ( 0.016 ) & 12.774 ( 0.020 ) \nl
       931.79800& 11.771 ( 0.054 ) & 12.366 ( 0.017 ) & 12.465 ( 0.013 ) &
 12.436 ( 0.015 ) & 12.768 ( 0.020 ) \nl
       932.74700& 11.825 ( 0.016 ) & 12.375 ( 0.020 ) & 12.465 ( 0.017 ) &
 12.443 ( 0.017 ) & 12.843 ( 0.026 ) \nl
       932.74700& 11.819 ( 0.019 ) & 12.378 ( 0.019 ) & 12.467 ( 0.017 ) &
 12.454 ( 0.013 ) & 12.816 ( 0.031 ) \nl
       934.76300& 11.993 ( 0.017 ) & 12.448 ( 0.023 ) & 12.499 ( 0.018 ) &
 12.441 ( 0.017 ) & 12.886 ( 0.025 ) \nl
       934.76300& 11.981 ( 0.012 ) & 12.451 ( 0.023 ) & 12.506 ( 0.013 ) &
 12.451 ( 0.019 ) & 12.878 ( 0.032 ) \nl
       936.67200& 12.118 ( 0.041 ) & 12.558 ( 0.027 ) & 12.547 ( 0.017 ) &
 12.529 ( 0.026 ) & 12.971 ( 0.022 ) \nl
       936.67200& 12.117 ( 0.052 ) & 12.560 ( 0.022 ) & 12.563 ( 0.015 ) &
 12.526 ( 0.018 ) & 12.966 ( 0.027 ) \nl
       937.81700& 12.234 ( 0.042 ) & 12.634 ( 0.024 ) & 12.590 ( 0.018 ) &
 12.579 ( 0.016 ) & 13.091 ( 0.038 ) \nl
       937.81700& 12.245 ( 0.023 ) & 12.633 ( 0.025 ) & 12.597 ( 0.024 ) &
 12.577 ( 0.022 ) & 13.084 ( 0.032 ) \nl
       939.75200& 12.453 ( 0.052 ) & 12.767 ( 0.047 ) & 12.686 ( 0.016 ) &
 12.715 ( 0.017 ) & 13.190 ( 0.020 ) \nl
       939.75300& 12.429 ( 0.034 ) & 12.778 ( 0.029 ) & 12.681 ( 0.013 ) &
 12.716 ( 0.018 ) & 13.199 ( 0.016 ) \nl
       949.70100& 13.799 ( 0.026 ) & 13.843 ( 0.022 ) & 13.286 ( 0.026 ) &
 13.249 ( 0.019 ) & 13.389 ( 0.029 ) \nl
       949.70100& 13.812 ( 0.016 ) & 13.837 ( 0.031 ) & 13.312 ( 0.023 ) &
 13.303 ( 0.020 ) & 13.400 ( 0.024 ) \nl
       951.74600& 14.116 ( 0.025 ) & 14.118 ( 0.021 ) & 13.405 ( 0.019 ) &
 13.294 ( 0.032 ) & 13.336 ( 0.030 ) \nl
       951.74600& 14.118 ( 0.028 ) & 14.126 ( 0.024 ) & 13.417 ( 0.016 ) &
 13.281 ( 0.013 ) & 13.337 ( 0.025 ) \nl
       953.70700& 14.416 ( 0.029 ) & 14.361 ( 0.020 ) & 13.524 ( 0.018 ) &
 13.308 ( 0.028 ) & 13.289 ( 0.028 ) \nl
       953.70700& 14.403 ( 0.055 ) & 14.369 ( 0.029 ) & 13.527 ( 0.018 ) &
 13.342 ( 0.023 ) & 13.289 ( 0.025 ) \nl
       956.65600& 14.787 ( 0.035 ) & 14.686 ( 0.031 ) & 13.708 ( 0.024 ) &
 13.368 ( 0.029 ) & 13.226 ( 0.020 ) \nl
       956.65700& 14.775 ( 0.045 ) & 14.714 ( 0.019 ) & 13.667 ( 0.032 ) &
 13.408 ( 0.025 ) & 13.267 ( 0.026 ) \nl
       960.69600& 15.130 ( 0.024 ) & 15.009 ( 0.019 ) & 13.892 ( 0.020 ) &
 13.514 ( 0.020 ) & 13.231 ( 0.018 ) \nl
       960.69700& 15.132 ( 0.018 ) & 15.018 ( 0.017 ) & 13.902 ( 0.024 ) &
 13.536 ( 0.040 ) & 13.235 ( 0.025 ) \nl
       963.64600& 15.243 ( 0.051 ) & 15.211 ( 0.017 ) & 14.063 ( 0.015 ) &
 13.724 ( 0.038 ) & 13.358 ( 0.024 ) \nl
       963.64600& 15.273 ( 0.039 ) & 15.209 ( 0.020 ) & 14.102 ( 0.051 ) &
 13.692 ( 0.015 ) & 13.365 ( 0.022 ) \nl
       980.68000& 15.773 ( 0.029 ) & 15.785 ( 0.034 ) & 14.859 ( 0.037 ) &
 14.526 ( 0.017 ) & 14.375 ( 0.017 ) \nl
       982.70100& 15.829 ( 0.046 ) & 15.778 ( 0.048 ) & 14.940 ( 0.050 ) &
 14.654 ( 0.033 ) & 14.476 ( 0.036 ) \nl
       987.69200& 15.877 ( 0.029 ) & 15.828 ( 0.013 ) & 14.965 ( 0.020 ) &
 14.756 ( 0.017 ) & 14.713 ( 0.018 ) \nl
       987.69300& ... & ... & 14.946 ( 0.017 ) &
 ... & ... \nl
       987.70100& ... & ... & 14.954 ( 0.018 ) &
 ... & ... \nl
       995.67900& 16.014 ( 0.032 ) & 15.942 ( 0.018 ) & 15.161 ( 0.019 ) &
 14.998 ( 0.016 ) & 15.080 ( 0.024 ) \nl
       1010.6530& 16.388 ( 0.035 ) & 16.138 ( 0.015 ) & 15.560 ( 0.012 ) &
 15.503 ( 0.011 ) & 15.691 ( 0.026 ) \nl
       1013.6570& 16.470 ( 0.033 ) & 16.197 ( 0.024 ) & 15.642 ( 0.023 ) &
 15.593 ( 0.019 ) & 15.806 ( 0.023 ) \nl
       1013.6630& ... & ... & 15.650 ( 0.018 ) &
 ... & ... \nl
       1013.6700& ... & ... & 15.639 ( 0.020 ) &
 ... & ... \nl
       1024.6430& 16.648 ( 0.043 ) & 16.346 ( 0.032 ) & 15.913 ( 0.019 ) &
 15.941 ( 0.027 ) & 16.214 ( 0.031 ) \nl
       1137.0280& 19.328 ( 0.142 ) & 18.090 ( 0.019 ) & 18.028 ( 0.077 ) &
 18.625 ( 0.145 ) & 18.587 ( 0.044 ) \nl
       1195.9130& 20.455 ( 0.336 ) & 18.956 ( 0.031 ) & 18.833 ( 0.087 ) &
 19.599 ( 0.144 ) & ... \nl
       1216.9110& ... & 19.299 ( 0.032 ) & 19.291 ( 0.035 ) &
 ... & ... \nl
       1224.8590& ... & 19.433 ( 0.067 ) & 19.427 ( 0.039 ) &
 20.146 ( 0.230 ) & 19.998 ( 0.167 ) \nl
       1275.8010& ... & 20.013 ( 0.074 ) & 20.140 ( 0.131 ) &
 ... & ... \nl
\enddata 
\tablenotetext{a}{MJD is the Julian date minus 2450000.}
\end{deluxetable}

\clearpage

  A useful crosscheck of the consistency of our photometry would be to compare
the magnitudes of comparison stars in the SN fields as determined with the
$HST$ data and with the ground-based data.  However, the dramatic difference in
the sensitivity between the ground and $HST$ results in a lack of stars which
are well-detected but not saturated in either observations.  For our $HST$
images, stars with $V < 19.75$ mag and $I < 19.5$ mag are saturated, but stars
fainter than these limits are poorly detected (if at all) in the SN follow-up
images obtained from the ground.  In order to produce a comparison we combined
all of the deep follow-up imaging from the ground to increase the sensitivity.
Consequently, three stars in the NGC 3370 field (labeled as Stars 1a, 2a, and
3a) can be compared.  For these three stars the mean difference in the $V$ band
is $0.02 \pm 0.04$ mag (ground brighter), in good agreement.  In the $I$ band,
the only star which is not saturated is Star 3a, whose ground-based magnitude
is $0.04 \pm 0.17$ mag brighter than measured by $HST$, in good agreement but
not of useful precision.  No unsaturated stars in the NGC 3982 field are
visible in the $HST$ images.

   The optical light curves of the other ideal calibrators, SN 1990N and SN
1981B, are are taken from Lira et al. (1998) and from Buta \& Turner (1983),
respectively.

\section{Measuring the Hubble Constant}

\subsection{Light-Curve Fitting and Distance Calibrations}

   SNe~Ia are the most precise distance indicators known to delineate the
global Hubble flow.  Methods which utilize the relationship between luminosity,
light curves, and colors achieve relative individual distance precision of 7\% to
8\%.  With independent measurements of the distance to a few SN~Ia hosts from
Cepheids or other means, we can establish an absolute distance scale for SNe~Ia
and as a result, determine the Hubble constant.

   Here we employ two popular methods for measuring SN~Ia distances from their
light curves, the ``multicolor light-curve shape'' method (MLCS; Riess et al. 1996a)
and a template-fitting method ($\Delta m_{15}(B)$; Phillips 1993).  In both cases (and
equivalently applicable to other light-curve relations), we will use the
Cepheid-based distance measurements to the hosts of SN 1994ae and SN 1998aq to
set each method's fiducial luminosity (i.e., zero-point).

   The MLCS method has been revised (and hereafter referred to as MLCS2k2) by Jha 
(2002) and Jha, Riess, \& Kirshner(2005b) to include $U$-band light curves from Jha 
et al. (2005a), a more self-consistent treatment of extinction, and an improved
determination of the unreddened SN~Ia color.  In MLCS2k2 the light curves are
empirically described in any bandpass $X$ as a superposition of time-dependent
vectors in magnitudes:

$$\overrightarrow{m_X}(t-t_0)=\overrightarrow{M^0_X}+\mu_0+\overrightarrow{\eta_X}A^0_V+\overrightarrow{P_X}\Delta
+ \overrightarrow{Q_X}\Delta^2, $$

\noindent 
where $\overrightarrow{m_X}(t-t_0)$ is the apparent magnitude light curve since
a fiducial point in time $t_0$ (taken to be the time of $B$-band peak),
$\overrightarrow{M^0_X}$ is the absolute magnitude light curve of the fiducial
SN~Ia, $\mu_0$ is the true distance modulus, $A^0_V$ is the visual extinction
at $t_0$, $\eta_X$ contains the time-dependent reddening law (see Jha et
al. 2005b), $\Delta$ is the luminosity correction corresponding to
$M_V(t_0)-M^0_V(t_0)$, and $P_X$ and $Q_X$ are vectors describing the change in
light-curve shape with luminosity.  After determining the form of $P_X$ and
$Q_X$ from a Hubble-flow training set, the only free parameters for an
individual SN Ia are $t_0$, $\mu_0$, $\Delta$, and $A^0_V$.

   To determine the unreddened colors of the fiducial SN~Ia, MLCS2k2 uses the
observation that by $\sim$35 days after $B$ maximum, when SNe~Ia have entered
the nebular phase, their colors are uniform (empirically shown by Lira 1996;
Riess, Press, \& Kirshner 1996a).  Any variation from this color is assumed to
be due to reddening and intrinsic dispersion.  Thus, the distribution of
observed nebular colors can be fit with three parameters: a fiducial color, the
intrinsic dispersion of the Gaussian distribution of colors, and the scale of
an exponentially decaying distribution of reddenings (motivated by simulations
of host opacities from Hatano, Branch, \& Deaton 1998).  Using the derived,
unreddened nebular colors, all SNe~Ia are dereddened and utilized to derive the
$P_X$ and $Q_X$ vectors.  The end result is a determination that the fiducial
SN~Ia (which for MLCS2k2 has $\Delta m_{15}(B) = 1.07$ mag) has unreddened
colors at $B$-band maximum of $U-B = -0.40$, $B-V = -0.11$, $V-R = 0.00$, and 
$V-I = -0.30$ mag.

   For an individual SN~Ia, the time-dependent luminosity is described by
$\overrightarrow{M^0_X}+\overrightarrow{P_X}\Delta +
\overrightarrow{Q_X}\Delta^2$, where the sum of the second and third terms
represents the difference or correction to the luminosity from the fiducial to
the individual SN~Ia.  The zero-point or distance scale is set by a
determination of $\overrightarrow{M^0_X}$ at any phase (e.g., at peak).

   Here we measure the fiducial luminosity by using the Cepheid calibrations of
the distances to the hosts of SN 1994ae and SN 1998aq (now treated as the {\it
true} distances) and evaluating
$$ \overrightarrow{M^0_X}=\overrightarrow{m_X}(t-t_0)-\mu_{{\rm
Cepheid},0}-\overrightarrow{\eta_X}A^0_V - \overrightarrow{P_X}\Delta -
\overrightarrow{Q_X}\Delta^2, $$
\noindent 
where $\mu_{{\rm Cepheid},0}$ is the Cepheid-calibrated distance from the
previous section.

\clearpage

\begin{deluxetable}{cccccccc}
\tablecaption{MLCS2k2 Parameters}
\tablenum{12}
\tablehead{\colhead{SN} & \colhead{$\mu_{Cepheid,0}(\sigma^a)$} & \colhead{$M^0_V(t_0)$} & \colhead{$\sigma$} & \colhead{$\Delta(\sigma)$} & \colhead{$A^0_V(\sigma)$} & \colhead{$t_0^b$} & \colhead{$H_0^c$} \\ & (mag) & (mag) & (mag) & (mag) & (mag) & & }
\startdata
1994ae & $32.29 \pm 0.06$ & $-19.18$ & 0.12  & $-$0.33(0.03) & 0.30(0.03) &  9684.84(0.35) & 72 \nl
1998aq & $31.66 \pm 0.09$ & $-19.19$ & 0.12  & $-$0.05(0.02) & 0.05(0.03) &  50930.85(0.10) & 73 \nl
1990N  & $31.74 \pm 0.12$ & $-19.10$ & 0.12 & $-$0.32(0.06) & 0.42(0.05) & 8081.90(0.14) & 75 \nl
1981B  & $30.86 \pm 0.09$ & $-19.21$ & 0.12  & $-$0.19(0.03) & 0.52(0.05) & 44670.71(0.88) & 72 \nl
\enddata
\tablenotetext{a}{Statistical uncertainty from dereddened \PL fit and [O/H]
measurement only.}
\tablenotetext{b}{The Julian date minus 2400000.}
\tablenotetext{c}{The units of $H_0$ are km s$^{-1}$ Mpc$^{-1}$.}
\end{deluxetable}

\clearpage


  The MLCS2k2 parameters derived from the fits are given in Table 12.  The
parameter of greatest interest, $M_V^0(t_0)$, is $-19.18 \pm 0.13$ mag and
$-19.19 \pm 0.13$ mag for the new objects SN 1994ae and SN 1998aq,
respectively, where the uncertainty listed contains only the uncorrelated
(random) uncertainty as discussed in the next section.  A simple average of the
four ideal calibrators yields $M_V^0(t_0) = -19.17 \pm 0.07$ mag (the more
accurate weighted average computed in the next section yields the same result).

   Interestingly, the two ``corrections'' to the brightness of SN 1994ae and SN
1998aq, $\Delta$ and $A_V^0$, are nearly identical and offset each other (e.g.,
$\Delta + A_V^0 \approx 0$ mag), indicating that in the absence of any
corrections for intrinsic luminosity and extinction, the estimated distance
moduli would remain virtually identical.

   We have utilized the ``Gold'' sample of SNe~Ia from Riess et al. (2004) to
measure the global Hubble flow of SNe~Ia.  This full list is a complete,
``high-confidence'' set of 157 objects that includes all SNe~Ia not suffering
from one or more known sources of systematic uncertainty.  Reasons for
exclusion from the Gold sample are (1) an uncompelling classification or (2)
an incomplete or potentially inaccurate photometric record (including limited
light-curve sampling, low-precision color information, or non-CCD
measurements).  An additional reason for exclusion was a large measured
extinction ($A_V > 1$ mag), amplifying systematic errors from potentially
non-Galactic reddening laws.

   The relative distances of this sample were previously used to characterize
mass-energy components of the Universe.  However, the zero-point or distance
scale was set to $M_V^0(t_0) = -19.44$ mag in accordance with Cepheid analyses
by the STS group.  The values of $M_V^0(t_0)$ provided in Table 11 allow us to
set the luminosity of the SNe~Ia in the Gold sample to the Cepheid calibrations
presented here, resulting in a constant decrease of all distances in Riess et
al. (2004) by $M_V^0(t_0) - 19.44 = 0.27$ mag.

   The redshift range used to measure the global expansion rate has a
non-trivial impact on the inferred value.  Zehavi et al. (1998) first noted a
possible kinematic ripple in the Hubble diagram of SNe~Ia at $cz = 7000$ km
s$^{-1}$ representing a few percent change in the Hubble constant across this
feature.  To measure the true, global expansion rate free of undue influence by
flows and infalls, we have set the boundary beyond this feature.  To reduce any
``cosmological bias'' depending on the absence or presence of a 70\% fraction
of vacuum energy, we have set an upper limit of $z = 0.08$ to include the greatest
number of SNe~Ia while limiting such bias to less than 2\% of the Hubble
constant at an average redshift $z = 0.04$.  The 38 SNe Ia from the Gold 
sample in this redshift range are: SN1998eg, SN1994M, SN1993H, SN1999X, SN1999gp, SN1992ag, SN1992P, SN2000bk, SN1996C, SN1993ah, SN1994Q, SN1997dg, SN1990O, SN1999cc, SN1998cs, SN1991U, SN1996bl, SN1994T, SN1992bg, SN2000cf, SN1999ef, SN1990T, SN1992bl, SN1992bh, SN1992J, SN1995ac, SN1993ac, SN1990af, SN1993ag, SN1993O, SN1998dx, SN1991S, SN1992bk, SN1992au, SN1992bs, SN1993B, SN1992ae, and SN1992bp.  

   Following Jha et al. (1999), it is useful to define a
distance-scale-independent intercept of the Hubble-flow SNe~Ia, $a_V$, where
$a_V \equiv {\rm log}(cz) - 0.2m_V^0(t_0) = {\rm log}(H_0) -0.2 M_V^0(t_0) -
5$, which we find to be $0.697 \pm 0.005$ in units of 0.2 mag for the Gold
sample of Riess et al. (2004) at $0.023 < z< 0.080$.\footnote{The specific
value of $a_V$, though independent of distance scale, does depend on the
arbitrary choice of a fiducial SN~Ia, but this dependence is removed from the
determination of $H_0$.}  The value of $H_0$ for each ideal calibrator is then
given in Table 12 for its individual value of $M_V^0(t_0)$.

\subsection{Error Budget}

   To calculate the weighted average value of $H_0$ for all ideal calibrators
and its uncertainty, it is necessary to distinguish between errors which are
correlated or uncorrelated among the individual calibrators and thus may or may
not be reduced by averaging.  The propagation of uncertainty is tabulated in
Table 13 and described here.

   Uncorrelated, random errors are dominated by the dispersion of ideal
calibrators.  Although the statistical uncertainty of such well-sampled light
curves as these can be $<0.05$ mag, the dispersion appears to be dominated by
an intrinsic component (Riess et al. 1996a; Perlmutter et al. 1997; Phillips et
al. 1999; Tripp \& Branch 1999; Jha et al.  2005b).  Estimates of this
intrinsic component range from 0.10 to 0.17 mag.  Due to the highly
overdetermined calibration and light-curve sampling of these objects, we take
this dispersion per object to be 0.12 mag.  Additional random errors result
from the uncertainty of the dereddened \PL fit (intercept) and the measurement
of the host metallicities.  (Errors due to differences in the host extinction
laws from the standard Galactic reddening law are negligible because the
reddenings are small and such errors are mostly removed by dereddening both the
SNe and the Cepheids.) These random error sources can be added in quadrature
and are assumed to be independent for each individual collaborator.

   Each of the two $HST$ cameras, WFPC2 and ACS, are assumed to have
independent 0.03 mag uncertainties in their zero-points in both the $V$ and $I$
bands.  After dereddening the Cepheids, the resultant uncertainty is 0.09 mag
[i.e., $(\sigma_V^2(1-R)^2 + \sigma_I^2 R^2)^{1/2}$] for each camera.  For the
three calibrations obtained with WFPC2, the 0.09 mag error is correlated, but
is assumed to be uncorrelated with the 0.09 mag for the single ACS
calibration. We sum the camera-dependent error with the above individual errors
separately for the SNe calibrated with ACS and WFPC2.

   The weighted average value of $H_0$ for the four calibrators is computed in
proportion to their uncorrelated error components (i.e., the individual errors
determine the weights within the subtotal for each camera and then are combined
with the independent camera errors to reach the total as shown in Table 13).

   Correlated errors affecting all calibrators include an assumed 0.10 mag
uncertainty for the distance to the LMC (i.e., the consensus value of the STS
and SKP groups), an 0.05 mag uncertainty for the form (i.e., slope) of the \PL
relation, and the 0.025 mag uncertainty determined by the fit to the ridge line
of the 37 SNe~Ia defining the Hubble flow from the Gold sample.  The 25\%
systematic uncertainty in the metallicity correction from Sakai et al. (2004)
is negligible compared to the other systematic error sources. Together, these
errors provide an irreducible, systematic uncertainty of 0.12 mag.

\clearpage 
\begin{deluxetable}{lcccc} 
\footnotesize
\tablecaption{Error Budget for $H_0$ in Units of Magnitudes}
\tablenum{13}
\tablehead{\colhead{ }&\colhead{       }&\colhead{Random Error Sources}&\colhead{    }&\colhead{    }}
\startdata
\multicolumn{1}{l}{Source} & \multicolumn{1}{c}{1994ae(ACS)} & \multicolumn{1}{c}{1998aq(WFPC2)}& \multicolumn{1}{c}{1990N(WFPC2)} & \multicolumn{1}{c}{1981B(WFPC2)} \nl
\hline
Dereddened \PL fit & 0.05 & 0.08 & 0.09 & 0.07 \nl
MLCS2k2 $M_V^0(t_0)$ fit & 0.12 & 0.12 & 0.12 & 0.12 \nl
$[O/H]$ measurement & 0.03 & 0.03 & 0.08 & 0.05 \nl
\hline
R1. Individual total & 0.133 & 0.147 & 0.170 & 0.147 \nl
\multicolumn{5}{c}{} \nl
\multicolumn{1}{l}{HST camera zero-point} & \multicolumn{1}{c}{0.03(ACS)} & \multicolumn{3}{c}{$\longleftarrow$ 0.03(WFPC2) $\longrightarrow$} \nl
\multicolumn{1}{l}{R2. effect on dereddened \PL} & \multicolumn{1}{c}{0.086(ACS)} & \multicolumn{3}{c}{$\longleftarrow$ 0.086(WFPC2) $\longrightarrow$} \nl
\multicolumn{1}{l}{$R12=R1+R2$(ACS/WFPC2)} & \multicolumn{1}{c}{0.158(ACS)}&\multicolumn{3}{c}{$\longleftarrow$ 0.124(WFPC2) $\longrightarrow$} \nl
\multicolumn{2}{l}{Total Random Error =}& \multicolumn{3}{l}{0.098} \nl
\hline
\multicolumn{2}{c}{ }&\multicolumn{1}{c}{Systematic Error Sources} &\multicolumn{2}{c}{ }\nl
\hline
\multicolumn{2}{l}{LMC modulus} & \multicolumn{3}{l}{0.10} \nl
\multicolumn{2}{l}{Slope of \PL relation} & \multicolumn{3}{l}{0.05} \nl
\multicolumn{2}{l}{Hubble flow = $5\sigma_{a_V}$} & \multicolumn{3}{l}{0.025} \nl
\multicolumn{2}{l}{Total Systematic Error =}& \multicolumn{3}{l}{0.115} \nl
\hline
\multicolumn{2}{l}{Total Error =}& \multicolumn{3}{l}{0.150} \nl
\enddata
\end{deluxetable}
\clearpage

   The combined uncertainty for the four measurements is $1\sigma = 0.10$ mag
for the random component (now including the zero-points of the $HST$ cameras) and
$1\sigma = 0.12$ mag for the systematic component, for a total of 0.15 mag, or 8\%
in the value of $H_0$.  Weighting the average of the four calibrators (in Table
12) in proportion to the uncorrelated error components yields

$ H_0 = 73 \pm 4$ (statistical) $ \pm 5$ (systematic)$  = 73 \pm 6.4$(total) \ho.

   To test the sensitivity of our results to the adopted light-curve fitting
method, here we also applied the $\Delta m_{15}(B)$ fitting method as
calibrated by Suntzeff et al. (1999) and Phillips et al. (1999), and using the
fitted peak apparent magnitudes and $\Delta m_{15}(B)$ parameters given in
Table 14.\footnote{For each of the $UBVRI$-band light curves, we fit a closely
matching light-curve segment to the data in a small range near maximum
brightness ($-10 < t < 15$ days).  The best-fit peak apparent magnitudes are
then given in Table 14.  The high rate of data sampling allowed us to estimate
the values of $\Delta m_{15}(B)$ directly as given in Table 14.  Following
Phillips et al. (1999), the observed value of $\Delta m_{15}(B)$ for SN 1994ae
should be reduced by 0.01 mag (to 0.89 mag) to account for the light-curve
narrowing effect of $E(B-V) \approx 0.1$ mag.}  The result was a value of $H_0
= 71$ \ho\ (with a similar error budget as the one in Table 13), in good
agreement with the MLCS2k2 method.\footnote{From Phillips et al. (1999), we
use Equations (7) and (8) which express the unreddened pseudo-color
$B_{max}-V_{max}$ and $V_{max}-I_{max}$ of a SN~Ia; we find $E(B-V) = 0.11 \pm
0.04$ mag and $E(V-I) = 0.17 \pm 0.04$ mag for SN 1994ae, and $E(B-V) = -0.02
\pm 0.04$ mag and $E(V-I) = 0.07 \pm 0.04$ mag for SN 1998aq.  Using a standard
extinction law (Cardelli et al. 1989), the average extinction for
SN 1994ae and SN 1998aq is $A_V=0.26 \pm 0.10$ mag and $A_V=0.04 \pm 0.10$ mag,
respectively.  Using the expression from Suntzeff et al. (1999), ${\rm log}
H_0(V) = 0.2(M^V_{max}-0.672(\Delta m_{15}(B)-1.1)-0.633(\Delta
m_{15}(B)-1.1)^2 + 28.590)$, yields $H_0 = 70$ and 76 \ho, respectively. For SN
1990N and SN 1981B using the parameters from Suntzeff et al. (1999) and our
OGLE+10 \PL relation-based estimate of the Cepheid distances to the hosts
yields $H_0 = 69$ and 70 \ho, respectively.  The average value of the four
ideal calibrators is thus 71 \ho.} We have also attempted to reproduce the
methodology of the STS group and by their methods find an average value of $H_0 = 69$
\ho\ for SN 1994ae and SN 1998aq.\footnote{We can try to estimate the values of
$H_0$ that the STS collaboration would find from these two new calibrators (of
course, we cannot be sure what they will find until they analyze the data
presented here themselves).  To reproduce the STS prescription we take the $V$
at peak of 12.46 mag from Table 14, subtract the Saha et al. (2001) Cepheid
value of $\mu_0 = 31.72$ mag to yield $M_V^0(t_0) = -19.26$ mag.  Assuming STS
would find no significant corrections resulting from the low apparent reddening
and typical light-curve shape seen for SN 1998aq, we find $M_V^0(t_0)$ is
fainter than that in Parodi et al. (2000) by 0.27 mag, yielding $H_0$ which is
13.5\% larger than that of Parodi et al. (2000), or $H_0 = 69$ \ho.  The
difference between this and the value $H_0 = 60$ \ho\ from STS for SN 1998aq
results from the mis-estimate of the peak $V$ used by Saha et al. (2001), which
was based on a rough, amateur observation at a single phase and apparently 0.18
mag too bright as compared with the complete photometric record presented here.
For SN 1994ae, using the STS-method value of $\mu_w = 32.36$ mag (see \S 2.3)
and the similar lack of significant corrections results in $M_V^0 = 13.08-32.36
= -19.28$ mag for a similar result of $H_0 = 68$ \ho.}  Following the SKP
methodology yields an average value of $H_0 = 75$ \ho.  Although we conclude that
the inferred value of $H_0$ from this set of SN~Ia calibrators is not sensitive
to the light-curve shape method, we invite the application of other fitting
prescriptions to the data presented here.  The striking reduction in the
difference between the value of $H_0$ estimated by the prescriptions of
different groups results directly from the improvements in data quality as
enumerated in the next section.

\clearpage

\begin{deluxetable}{ccccccc}
\tablecaption{SN Observables$^a$}
\tablenum{14}
\tablehead{\colhead{SN}&\colhead{$U_{max}$ (mag)}&\colhead{$B_{max}~$
(mag)}&\colhead{$V_{max}$~(mag)}&\colhead{$R_{max}$~(mag)}&\colhead{$I_{max}$~(mag)}&\colhead{$\Delta m_{15}(B)$ }}
\startdata
SN 1994ae &  12.45(0.04) & 13.08(0.02) & 13.08(0.02) & 13.07(0.03) & 13.29(0.04) &  0.90(0.03) \nl
SN 1998aq &  11.70(0.03) & 12.36(0.02) & 12.46(0.02) & 12.43(0.02) & 12.72(0.02) &  1.05(0.03) \nl
\enddata
\tablenotetext{a}{The uncertainty is given in parentheses.}

\end{deluxetable}

\clearpage

\section{Discussion}

   The ``era of precision cosmology'' has been realized as an unprecedented
time when it has become possible to constrain each of the cosmological
parameters which characterize our Universe to an uncertainty of less than 10\%.
Arguably the most fundamental of the parameters and the one to which the
greatest effort has been applied is the Hubble constant.  Unfortunately, its
estimation has remained highly controversial, suffering from direct propagation
of errors along a long chain of ``nuisance parameters'' (e.g., the distance to
the LMC). SNe~Ia are the most precise, long-range distance indicator known and
their accurate calibration lies at the heart of estimating $H_0$.  The two most
active groups employing SNe~Ia to estimate $H_0$, the Key Project and the STS
collaboration, disagree by 20\% in their latest (perhaps final?) attempts.  We believe
the resolution of this difference is a prerequisite to adding $H_0$ to the
ledger of the well-calibrated parameters.

   In Table 15 we summarize the primary sources of differences between these
two groups and our own analysis that lead to the 20\% discrepancy in the value
of $H_0$.

\clearpage

\begin{deluxetable}{lcc}
\tablecaption{Sources of Differences in Estimation of $H_0$ using SNe~Ia}
\tablenum{15}
\tablehead{\colhead{Source}&\colhead{$\Delta H_0$ STS vs. this
paper}&\colhead{$\Delta H_0$ STS vs. SKP}}
 \startdata
Use of non-ideal calibrators$^a$ & $\downarrow$ 7\% & $\downarrow$ 1\% \nl
Form of \PL relation & $\downarrow$ 5\% & $\downarrow$ 7\% \nl
Measured extinction$^b$ & $\downarrow$ 5\% & $\downarrow$ 5\% \nl
Lack of metallicity correction & $\uparrow$ 4\% & $\uparrow$ 4\% \nl
WFPC2 ``long vs. short effect'' & $\downarrow$ 2.5\% & $\downarrow$ 2.5\% \nl
WFPC2 CTE correction & $\downarrow$ 2\% & $\downarrow$ 2\% \nl
$HST$ zero-points & $\downarrow$ 1.5\% & $\downarrow$ 3\% \nl
Use of pre-1980 SNe~Ia, Hubble flow$^c$ & $\downarrow$ 1\% & $\downarrow$ 1\% \nl
Strength of luminosity correction & $\downarrow$ 1\% & $\downarrow$ 1\% \nl
Cosmological model & $\downarrow$ 1\% & $\downarrow$ 1\% \nl
\hline
Total& $\downarrow$ 22\% & $\downarrow$ 20\% \nl 
\enddata
\tablenotetext{a}{Differences in $M_V^0$ from Saha et al. 2001 for 1981B and
1990N vs. the rest.}
\tablenotetext{b}{STS use a redder intrinsic color resulting in less
extinction, most affecting SNe~Ia in Cepheid hosts; estimated from the
difference in $A_V$ for SN 1981B + SN 1990N and the Gold sample.}
\tablenotetext{c}{Pre-1980 vs. post-1980 SNe~Ia average $M_V^0$ from Parodi et
al. is 0.09 mag, Pre-1980 SNe Ia are 25\% of sample.}
\end{deluxetable}

\clearpage

   We conclude that much of the discrepancy is caused by the challenges of
wringing measurements from low-quality or problematic data --- photographic
photometry, highly extinguished or extreme SNe~Ia, poorly sampled light curves,
and cameras with photometric anomalies such as the ``long vs. short effect''
(Holtzman et al. 1995) and charge-transfer efficiency (CTE; e.g., Whitmore,
Heyer, \& Casertano 1999).  We estimate that together, these effects account
for fully $\sim$17\% of the 22\% difference in $H_0$ measurements in this paper and a similar fraction for the SKP
(see Table 15).  Debatable differences like the strength of the light curve
shape-luminosity relation impact $H_0$ by only 1\% (see Saha et al. 2001
for comparison).
The most expedient way to overcome the challenges
of poor data is to supercede the analysis with higher-quality data.  Using the
ideal calibrators presented here, we conclude that the remaining differences in
the two teams' approaches would appear to affect the Hubble constant by only
$\sim$ 5\%.\footnote{Most of this difference arises from the STS
collaboration's use of the \PL relation from the MF91 Cepheids, but the
superior sampling of the OGLE Cepheids (Udalski et al. 1999) and photometric
testing by Sebo et al. (2002) now argue in favor of an OGLE-based \PL relation,
closer to what has been adopted by SKP or by Thim et al. (2003).}

\subsection{More Accurate $H_0$ through Better SN~Ia Data}

   The renaissance in the use of SNe~Ia for precision cosmology has been fueled
by highly precise and accurate photometry.  Large samples ($> 10$ objects) of
CCD photometry, frequently using galaxy subtraction techniques and consistent
Landolt (1992) standards, have been published by the Cal\'{a}n/Tololo Survey
(Hamuy et al. 1996b), the CfA Survey I (Riess et al. 1999a), and the CfA Survey
II (Jha et al. 2005a). These datasets have individual photometric measurements
with absolute zero-points known to better than 5\% and lead to Hubble diagrams
with dispersions of less than 0.2 mag.

  Photographic observations obtained at the Asiago Observatory (and elsewhere)
of SNe~Ia in the 1950s through the early 1980s, while valuable for elucidating
the general properties of SNe~Ia, suffer from a significant lack of accuracy.
Despite recent efforts to recalibrate their standard-star sequences digitally
(Patat et al. 1997), these light curves still suffer from the original
inaccuracy in their SN magnitude estimates.  As described by Patat et
al. (1997), ``SN magnitudes, especially for objects lying on the luminous
background of their parent galaxies, were often obtained by eye comparison,'' a
statement which {\it still} describes the current state of photometry for these
supernovae even after adjustments to the comparison stars. Patat et al. (1997)
estimate that the errors from the by-eye estimates are typically $\sim$0.1 mag
(or 3 to 4 times this value after reddening correction), with noted
discrepancies of up to 0.6 mag from expected behavior at late times due to
``the poor contrast of the supernova against the parent galaxy background.'' The mean
$B-V$ at peak of the sample has changed by 0.06 mag (again, 3 to 4 times this
for reddening-free distance estimates) due only to corrections to the
comparison stars.  A similar sized, systematic error in the mean may still
plague the SN magnitudes based on the original by-eye estimations from the
photographic plates.  Additional errors may result from the inability to apply
``S-corrections'' (Suntzeff et al. 1999) to correct for the disparity between
Johnson/Cousins passbands and photographic emulsion sensitivities.\footnote{Anecdotally, 
we note that one photographic observation in the $B$ band of SN
1994ae by Tsvetkov \& Pavlyuk (1997) was obtained contemporaneously with our
observations on 9 Dec. 1994, for which Tsvetkov \& Pavlyuk (1997)
report the SN to be 0.25 mag brighter than our own CCD measurement.}

   Although it is regretable to discard data and further deplete the already
sparse, published sample of nearby SNe~Ia, it is difficult if not improbable to
extract accurate estimates of distances, free of $\sim$ 5--10\% systematic
errors, from such data.  Despite heroic efforts to recalibrate old
photographically observed SNe~Ia whose hosts are near enough to observe
Cepheids with WFPC2 (e.g., Schaefer 1994, 1995, 1996), in the interest of
accuracy and resolving the debate over the value of $H_0$ we discard (and
advocate others to discard) the use of SN 1895B, SN 1937C, SN 1960F, and SN
1974G.  Moreover, we think it is unwise to use such objects to help define the
Hubble flow or the luminosity vs. light-curve shape relation as done by the STS
collaboration (Parodi et al. 2000) and others (e.g., Rowan-Robinson 2002).  The
photometry of SN 1972E, obtained with a photoelectric photometer by
Kirshner et al. (1973), is likely better than photographic photometry, but
still was not obtained in the same way as for the SNe~Ia that now define the
Hubble flow.  As noted in Table 15, the use of pre-1980 SN~Ia data (both in the Hubble flow and among calibrators ) alone
accounts for 8\% of the change in $H_0$ from the STS group.

   An additional challenge is posed by SNe~Ia with high reddening (e.g.,
$E(B-V) > 0.3$ mag).  Although the {\it color excess} can be measured with
similar precision as it is for SNe~Ia having low reddening, inaccuracy may
arise in applying an estimate of the total-to-differential absorption ratio
[$R_V = A_V/E(B-V)$] for the SN host.  Galactic-type values are usually assumed
and tests show these are accurate to $\sim$ 25\% in the mean. [Likewise,
empirical estimation, such as by Tripp \& Branch (1999) and Parodi et
al. (2000), of the magnitude-color dependence are applicable in the mean.]
However, a 25\% uncertainty in $R_V$ will produce an error of 0.2--0.3 mag in
the distance modulus for $E(B-V) \approx 0.3$. Large observed extinction may
also arise from non-equilibrium processes (e.g., stellar mass loss) and be
characterized by unexpected values of $R_V$.  Riess et al. (2004) and Tonry et al. 
(2003) relegated SNe~Ia with large measured reddening to a ``low-confidence'' sample.
SN 1989B and SN 1998bu each suffer from large ($E(B-V) > 0.3$ mag) reddening
and therefore are not ideal calibrators.

   Differences in reddening-correction methods may also lead to dispute in the
value of $H_0$ if the average reddening of Hubble-flow SNe~Ia diverges from that
of calibrators.  In fact, this circumstance is likely because calibrators must
be chosen from hosts with ongoing massive-star formation (i.e., Cepheids) and
its accompanying dust.  (This circumstance even occurs to some degree for SN
1990N and SN 1981B, where the average $A_V$ for these two calibrators exceeds
that of SN 1994ae and SN 1998aq by 0.2 mag.)

   Also problematic for resolution of the $H_0$ debate are SNe~Ia whose
light-curve shapes or spectroscopic properties lie far from the mean or median
of SNe~Ia.  The relatively large distance correction implied by the extreme
nature of these SNe amplifies differences in the calibration of the correction
techniques.  (To those who derive a relatively ``weak'' correction, others
appear to overcorrect these objects, and vice versa.)  The second problem is
that the relatively low frequency of extreme SNe~Ia, especially in the Hubble
flow, reduces the objects available to define the calibration of the correction
techniques at the extreme.  Because our goal in this study is to resolve the 
disagreement over the value of $H_0$ determined from SNe~Ia,
we avoid the use of such extreme or peculiar SNe~Ia (and their added
controversy) as SN 1991T and SN 1999by.

   What remains are the ``ideal'' calibrators, objects having typical
light-curve shape, exquisite photometry, well-sampled light curves, and low
reddening.  This set includes SN 1990N, SN 1981B, SN 1994ae, and SN 1998aq.
The accuracy and precision of these objects offers the means to reduce the
long-standing disagreement over the Hubble constant determined from SNe~Ia.
We encourage the reanalysis of these objects and the similarly high-confidence
SNe~Ia in the Hubble flow by others to examine the size of the remaining
differences.

\subsection{WFPC2 vs. ACS}

   In the $HST$ Key Project's final report (Freedman et al. 2001), the
uncertainty in the photometric calibration of WFPC2 is listed as the
second-largest remaining source of error (after the distance to the LMC),
contributing a systematic error of 0.07--0.09 mag (Gibson et al. 2000; F01).
Cepheid calibrations with ACS on $HST$ offer several advantages for reducing this
source of uncertainty.

  The estimated dimming of Cepheids observed by WFPC2 due to imperfect CTE
reached 0.05 mag after 8 years of radiation damage in orbit by the time NGC
3982 was observed (Whitmore et al. 1999; Dolphin 2000).  However, significant
disagreement exists in the application of CTE corrections, with the STS
collaboration applying none to their DoPhot photometry (Saha et al. 2001) and
Stetson et al. (1998) utilizing only a time-independent component in their CTE
corrections.  In contrast, ACS was a new, undamaged camera at the time of our
observations with greater initial CTE than WFPC2.  Calibrations by Riess \&
Mack (2004) indicate that the typical corrections to Cepheid photometry of NGC
3370, obtained 1 year after ACS installation, is no more than a few millimag,
small enough to be insignificant and more importantly, uncontroversial.

   In addition, there has been disagreement over the existence of a 0.05 mag
photometric anomaly on WFPC2 known as the ``long vs. short effect.''  The
nature of the effect, first posited by Holtzman et al. (1995), is a net decrease
in the flux of short exposures relative to long exposures by 5\%, although more
recent analysis by Whitmore et al. (1999) indicates that it may result from inclusion
of nearby stars in the annuli used to quantify local background levels in dense
calibration clusters.  Generally, the STS collaboration has explicitly
accounted for this effect while the SKP has not.  For ACS photometry, no such
anomaly is known to exist (indeed, the linearity of the ACS detectors has been
shown to exquisite precision and accuracy; Gilliland et al. 2004).

   Two more advantages of ACS over WFPC2 are its field-of-view and pixel
sampling, both a factor of two greater for ACS.  The larger field-of-view for
ACS approximately doubles the number of Cepheids found in host galaxies of the
typical angular size encountered in past work.  The greater sampling (ACS
samples the PSF near the Nyquist limit; WFPC2/WFC undersamples the PSF by a
factor of two) improves the precision of individual photometry measurements and
can reduce the impact of crowding.  Together, these improvements reduce the
uncertainty for NGC 3370 (see Table 12) by $\sim$50\% over other WFPC2-based
calibrations.

   The calibration of photometric zero-points for each camera are expected to be
of similar quality.  Each depends on either knowledge of the
wavelength-dependent throughput or empirically determined color terms.
However, the uncertainty in the zero-points of each camera are relatively
independent, so additional calibrations from ACS should reduce the overall
systematic zero-point uncertainty.

   Yet, the biggest advantage of ACS is the extension in distance over WFPC2 to
which Cepheids can be measured in an economical amount of observing time.  This
added distance increases the sparse sample of SNe~Ia whose luminosities can be
calibrated with {\it HST} and improves the quality of the choices.

   We expect the largest source of uncertainty to now be (if not previously)
the distance estimate to the LMC.  However, further work on the geometric maser
distance to NGC 4258 (Herrnstein et al. 1999) and forthcoming observations of
its Cepheids with ACS may remove this uncertainty as well.  An added advantage
of NGC 4258 (besides its accurate distance) is the similarity of its
metallicity to the hosts of the ideal SN~Ia calibrators, reducing the reliance
on the accuracy of the metallicity corrections.

   Here we have realized a significant improvement by updating the calibration
of the Hubble constant from SNe~Ia by using recent SNe~Ia and a new $HST$
camera.  To continue to make progress in this field the trend of replacing
problematic data with modern data should continue.  As this occurs, we expect
the controversy over the value of $H_0$ as determined from SNe~Ia to gradually
abate.  If the disagreement over the value of $H_0$ by groups using the same
technique and the same data is taken as a measure of the uncertainty of its
value, we believe that uncertainty can be reduced by a factor of two or more in
the near future.

\bigskip 
\medskip 

   We wish to thank the Hubble Heritage team for help in producing a beautiful
image of NGC 3370.
Financial support for this work was provided by NASA through program GO-9353
from the Space Telescope Science Institute, which is operated by AURA, Inc.,
under NASA contract NAS 5-26555. Additional funding was provided by the
National Science Foundation for support of supernova research at Harvard
University (AST--0205808) and at the University of California, Berkeley
(AST--0307894).  The Berkeley Automatic Imaging Telescope at Leuschner
Observatory was funded by the NSF (AST--8957063) and the University of
California.  Some of the data presented herein were obtained at the W. M. Keck
Observatory, which is operated as a scientific partnership among the California
Institute of Technology, the University of California, and NASA; the
Observatory was made possible by the generous financial support of the
W. M. Keck Foundation. A.V.F. is grateful for a Miller Research Professorship
at U.C. Berkeley, during which part of this work was completed.

\vfill

\eject

\clearpage

\begin{deluxetable}{ccccccc}
\footnotesize
\tablenum{16}
\tablecaption{Photometry of Variables}
\tablehead{\colhead{ }&\colhead{ }&\colhead{ }&\colhead{ }&\colhead{ }&\colhead{ }&\colhead{ }}
\startdata
\hline
 MJD$^a$ & 71916 & 25970 & 8172. & 15128 & 29744 & 77676 \nl
\hline
\multicolumn{7}{c}{$F555W$} \nl
\hline
       52729.512& 26.76  0.05 & 26.48  0.03 & 26.30  0.03 & 27.72  0.14 &
 26.19  0.03 & 26.92  0.04 \nl
       52737.582& 26.62  0.05 & 27.13  0.03 & 26.19  0.03 & 29.12  0.36 &
 26.14  0.03 & 27.58  0.07 \nl
       52745.387& 26.97  0.06 & 26.20  0.02 & 26.43  0.04 & 28.40  0.20 &
 26.32  0.02 & 27.12  0.04 \nl
       52750.922& 27.13  0.06 & 26.46  0.03 & 26.80  0.06 & 26.50  0.03 &
 26.45  0.03 & 27.84  0.07 \nl
       52756.125& 26.67  0.05 & 26.59  0.03 & 26.80  0.05 & 26.28  0.02 &
 26.49  0.04 & 27.56  0.08 \nl
       52758.055& 26.08  0.02 & 26.83  0.04 & 26.98  0.06 & 26.37  0.04 &
 26.87  0.04 & 26.76  0.04 \nl
       52760.129& 26.11  0.03 & 26.95  0.04 & 27.16  0.08 & 26.62  0.03 &
 26.94  0.03 & 26.80  0.04 \nl
       52762.059& 26.07  0.03 & 26.98  0.05 & 27.03  0.06 & 26.55  0.05 &
 26.83  0.04 & 26.99  0.05 \nl
       52766.133& 26.12  0.03 & 26.95  0.05 & 26.94  0.06 & 26.75  0.04 &
 26.86  0.03 & 27.02  0.04 \nl
       52768.414& 26.31  0.04 & 27.23  0.05 & 26.05  0.03 & 27.14  0.07 &
 26.41  0.02 & 27.55  0.06 \nl
       52774.402& 26.24  0.04 & 27.12  0.04 & 25.86  0.03 & 27.35  0.11 &
 26.21  0.03 & 27.46  0.06 \nl
       52781.418& 26.45  0.03 & 26.19  0.03 & 26.11  0.03 & 27.21  0.07 &
 26.05  0.02 & 27.56  0.07 \nl
\hline
\multicolumn{7}{c}{$F814W$} \nl
\hline
       52729.645& 25.45  0.02 & 26.03  0.03 & 25.20  0.02 & 26.86  0.13 &
 25.32  0.02 & 26.69  0.06 \nl
       52745.516& 25.90  0.04 & 25.56  0.03 & 25.49  0.03 & 25.61  0.04 &
 25.41  0.02 & 26.52  0.05 \nl
       52760.340& 25.21  0.02 & 25.95  0.05 & 25.70  0.04 & 25.40  0.04 &
 25.78  0.02 & 26.12  0.04 \nl
       52768.547& 25.33  0.02 & 26.06  0.04 & 25.20  0.03 & 25.49  0.04 &
 25.34  0.02 & 26.57  0.06 \nl
       52781.469& 25.48  0.02 & 25.56  0.03 & 25.23  0.03 & 25.93  0.05 &
 25.27  0.02 & 26.03  0.04 \nl
\hline
 \enddata
\tablenotetext{a}{MJD is the Julian date minus 2400000.}
\end{deluxetable}

\clearpage

\vfill

\eject

Figure Captions

Figure 1: MLCS2k2 SN~Ia Hubble diagram using the Gold sample from Riess et
al. (2004).  $m^0_V(rest)$ is the peak apparent magnitude scaled to the
luminosity of the fiducial ($\Delta=0$) SN~Ia, corrected for reddening, and
transformed to the rest-frame $V$ band.  Conversion from apparent magnitudes to
distance requires knowledge of the distance scale or peak luminosity of the
fiducial SN~Ia.  Overplotted is the best fit for a flat cosmology ($\Omega_M =
0.29$, $\Omega_\Lambda = 0.71$).

Figure 2: Color images of a $3' \times 3'$ region of the field of NGC 3370 and
the SN~Ia 1994ae.  The left panel was produced from $B$, $V$, and $I$-band
images obtained at the FLWO 1.2-m telescope in December 1994, when SN 1994ae
was bright but well past maximum.  The right panel is the same region produced
from a stack of $F435W,$ $F555W,$ and $F814W$ images obtained with ACS on $HST$
in 2003.

Figure 3: Color images of a $3' \times 3'$ region of the field of NGC 3982 and
the SN~Ia 1998aq.  The left panel was produced from $B$, $V$, and $I$-band
images obtained at the FLWO 1.2-m telescope in 1998 when SN 1998aq was near
maximum brightness.  The right panel is the same region produced from a stack
of $F435W,$ $F555W,$ and $F814W$ images obtained with WFPC2 on $HST$ in 2001.

Figure 4: Hertzsprung-Russell diagram of 250 stars observed from the ground in
the cluster 47~Tuc and used to define a photometric transformation between ACS
passbands and the Johnson/Cousins system as calibrated by Landolt (1992).  
These stars were hand-selected as those which are well-detected in
both the ACS images and ground-based images from Stetson (2000).  Each of the
ground-based points is based on 25 to 37 independent measurements, with a final
photometric uncertainty of 2--3 millimag.

Figure 5: $F555W$ and $F814W$ light curves of Cepheids in NGC 3370.  For each
of the Cepheid candidates listed by identification number in Table 4, the
photometry is fit to the period-specific light-curve template.

Figure 6: Color-magnitude diagram of all stars in the master catalog for NGC
3370.  Large solid symbols show the positions of the Cepheids identified in NGC 3370, located as expected in the instability strip between the blue giants and  red giants (post-main-sequence stars).  
Figure 7: $V$-band and $I$-band period-luminosity relation for Cepheids
identified by one of us (P.B.S.) in NGC 3370. The \PL relation shown is from
the OGLE+10 Cepheids (Thim et al. 2003) and the observed dispersion is 0.31 and
0.24 mag for the $V$ and $I$ bands, respectively.  An approximate
instability-strip width of 0.35 mag is shown for comparison.

Figure 8: The $V$-band (upper), $I$-band (middle), and reddening-free (bottom)
distance modulus as a function of the lower period (magnitude limit) cutoff
used.  As shown, the distance moduli are relatively insensitive to the cutoff
due to the strong detectability of all Cepheids with ACS at the faint end.

Figure 9: Oxygen-to-hydrogen ratios of H~II regions in the host galaxies of two
SNe~Ia, NGC$\,$3370 and NGC$\,$3982.  Following the method of Zaritsky et al. 
(1994), emission-line intensity ratios of [O~II]
$\lambda\lambda$3726, 3729, [O~III] $\lambda\lambda$4959, 5007, and H$\beta$
were used to derive values of 12 + log(O/H) for H~II regions.  A linear
gradient was fit to determine the value of the metallicity correction at the
characteristic radial position of the observed Cepheids, $50''$. The
uncertainties shown are purely statistical; the dominant source of variance is
clearly systematic.

Figure 10: Ground-based comparison stars in the field of NGC 3370 used to
calibrate the light curves of SN 1994ae.  Images from the 1.2-m telescope 
at the  FLWO.

Figure 11: Ground-based comparison stars in the field of NGC 3982 used to
calibrate the light curves of SN 1998aq.  Image from the 1.2-m telescope at 
the FLWO.

Figure 12: MLCS2k2 fits to the $UBVRI$-band data for SN 1994ae and SN 1998aq.

\vfill 
\eject

{\bf References}

\refitem
Ayani, K., \& Yamaoka, H. 1998, IAU Circ. No. 6878

\refitem
Bennett, C., et al. 2003, ApJS, 148, 1

\refitem
Boisseau, J. R., \& Wheeler, J. C. 1991, AJ, 101, 1281

\refitem
Branch, D., Fisher, A., \& Nugent, P. 1993, AJ, 106, 2383

\refitem
Buta, R., \& Turner, A. 1983, PASP, 95, 72

\refitem
Cardelli, J. A., Clayton, G. C., \& Mathis, J. S. 1989, ApJ, 345, 245

\refitem
de Vaucouleurs, G., de Vaucouleurs, A., Corwin, H. G., Jr., Buta, R. J.,
   Paturel, G., \& Fouqu\'{e}, R. 1991, Third Reference Catalogue of 
   Bright Galaxies (New York: Springer)

\refitem
Dolphin, A. E. 2000, PASP, 112, 1383

\refitem
Ferrarese, L.,  et al. 2000, PASP, 112, 177

\refitem
Filippenko, A. V. 1997, ARA\&A, 35, 309

\refitem
Freedman, W., et al. 1994, ApJ, 427, 628

\refitem
Freedman, W., et al. 2001, ApJ, 553, 47

\refitem
Fruchter, A., \& Hook, R. 1997, SPIE, 3164, 120F

\refitem
Garcia, A. M. 1993, A\&AS, 100, 47 

\refitem
Gibson, B. K., et al. 2000, ApJ, 529, 723

\refitem
Gilliland, R., et al. 2004, ACS ISR 2004-01

\refitem
Hamuy, M., Phillips, M. M., Maza, J., Suntzeff, N. B.,
   Schommer, R. A., \& Aviles, R. 1995, AJ, 109, 1

\refitem
Hamuy, M., Phillips, M. M., Maza, J., Suntzeff, N. B.,
   Schommer, R. A., \& Aviles, R. 1996a, AJ, 112, 2398

\refitem
Hamuy, M., Phillips, M. M., Maza, J., Suntzeff, N. B.,
   Schommer, R. A., \& Aviles, R. 1996b, AJ, 112, 2408

\refitem
Hatano, K., Branch, D., \& Deaton, J. 1998, ApJ, 502, 177

\refitem
Herrnstein, J. R., et al. 1999, Nature, 400, 539

\refitem
Hill, R. J., et al., 1998, ApJ, 496, 648 

\refitem
Ho, W. C. G., Van Dyk, S. D., Peng, C. Y., Filippenko, A. V.,
  Leonard, D. C., Matheson, T., Treffers, R. R., \& Richmond, M. W. 
  2001, PASP, 113, 1349  

\refitem
Holtzman, J. A., Burrows, C. J., Casertano, S., Hester, J. J., Trauger, J. T.,
    Watson, A. M., \& Worthey, G. 1995, PASP, 107, 1065

\refitem
Hurst, G. M., Armstrong, M., \& Arbour, R. 1998, IAU Circ. No. 6875

\refitem
Iijima, T., Cappellaro, E., \& Turatto, M. 1994, IAU Circ. No. 6108

\refitem
Jha, S. 2002, Ph.D. thesis, Harvard University

\refitem
Jha, S., et al. 1999, ApJS, 125, 73

\refitem
Jha, S., et al. 2005a, AJ, submitted

\refitem
Jha, S., Riess, A. G., \& Kirshner, R. P. 2005b, in prep.

\refitem
Kennicutt, R., et al. 1998, ApJ, 498, 181

\refitem
Kirshner, R. P., Willner, S. P., Becklin, E. E., Neugebauer, G., \& Oke, 
   J. B. 1973, ApJ, 180, L97

\refitem
Koekemoer, A., et al. 2005, in prep 

\refitem
Landolt, A. 1992, AJ, 104, 340

\refitem
Leonard, D. C., Kanbur, S. M., Ngeow, C. C., \& Tanvir, N. R. 2003, 
   ApJ, 594, 247

\refitem
Lira, P. 1995, Masters thesis, University of Chile 

\refitem
Lira, P., et al. 1998, AJ, 115, 234 (Erratum: 1998, 116, 1006)

\refitem
Madore, B. F. 1982, ApJ, 253, 575

\refitem
Madore, B. F., \& Freedman, W. L. 1991, PASP, 103, 933

\refitem
Matheson, T., et al. 2005, in prep.

\refitem
Oke, J. B., et al. 1995, PASP, 107, 375

\refitem
Osterbrock, D. E. 1989, Astrophysics of Gaseous Nebulae and Active
   Galactic Nuclei (Mill Valley: Univ. Science Books)

\refitem
Parodi, B. R., et al. 2000, ApJ, 540, 634

\refitem
Patat, F., Barbon, R., Cappellaro, E., \& Turatto, M. 1997, 
    A\&A, 317, 423

\refitem
Perlmutter, S., et al. 1997, ApJ, 483, 565

\refitem
Perlmutter, S., et al. 1999, ApJ, 517, 565

\refitem 
Phillips, M. M. 1993, ApJ, 413, L105

\refitem
Phillips, M. M., et al. 1999, AJ, 118, 1766

\refitem
Richmond, M. W., Treffers, R. R., \& Filippenko, A. V. 1993, PASP,
   105, 1164

\refitem
Riess, A. G., \& Mack, 2004, STScI ACS Instrument Science Report 2004-06

\refitem
Riess, A. G., Press, W. H., \& Kirshner, R. P. 1995, ApJ,
  438, L17

\refitem
Riess, A. G., Press, W. H., \& Kirshner, R. P. 1996a, ApJ,
    473, 88

\refitem
Riess, A. G., Press, W. H., \& Kirshner, R. P. 1996b, ApJ,
   473, 588

\refitem
Riess, A. G., et al. 1998, AJ, 116, 1009

\refitem
Riess, A. G., et al. 1999a, AJ, 117, 707

\refitem
Riess, A. G., et al. 1999b, AJ, 118, 2675

\refitem
Riess, A. G., et al. 2004, ApJ, 607, 665

\refitem
Rowan-Robinson, M. 2002, MNRAS, 332, 352

\refitem
Saha, A., et al. 1996, ApJ, 466, 55

\refitem
Saha, A., et al. 1997, ApJ, 486, 1

\refitem
Saha, A., et al. 2001, ApJ, 562, 314 

\refitem
Sakai, S., Ferrarese, L., Kennicutt, R. C., \& Saha, A. 2004, 608, 42

\refitem
Schaefer, B. E. 1994, ApJ, 426, 493

\refitem
Schaefer, B. E. 1995, ApJ, 447, L13

\refitem
Schaefer, B. E. 1996, ApJ, 460, L19

\refitem
Sebo, K.M., et al. 2002, ApJS, 142, 71

\refitem
Sirianni, M., et al. 2005, submitted, PASP

\refitem
Spergel, D. N., et al. 2003, ApJS, 148, 175

\refitem
Stetson, P. B. 1987, PASP, 99, 191

\refitem
Stetson, P. B. 2000, PASP, 112, 925

\refitem
Stetson, P. B. 1993, in {\it Calibrating the Hubble Space Telescope}, in J.C. Blades and S.J. Osmer, ed(s)., Calibrating Hubble Space Telescope, Space Telescope Science Institute, Baltimore, pg 89, ``Reduction of WFC Images with DAOPHOT III''

\refitem
Stetson, P. B., \& Gibson, B. K. 2001, MNRAS, 328, L1

\refitem
Stetson, P. B., et al. 1998, ApJ, 508, 491

\refitem
Suntzeff, N., et al. 1999, AJ, 117, 1175

\refitem
Tammann, G. A., \& Reindl, B. 2002, Ap\&SS, 280, 165

\refitem
Tanvir, N. R. 1997, in ``The Extragalactic Distance Scale,'' ed. M. Livio
    (Cambridge: Cambridge Univ. Press), 91

\refitem
Thim, F., Tammann, G. A., Saha, A., Dolphin, A., Sandage, A., Tolstoy, E., \&
  Labhardt, L. 2003, ApJ, 590, 256

\refitem
Tonry, J. L., et al. 2003, ApJ, 594, 1

\refitem
Treffers, R. R., Filippenko, A. V., Van Dyk, S. D., Paik, Y., \&
     Richmond, M. W. 1995, in ``Robotic Telescopes: Current Capabilities,
     Present Developments, and Future Prospects for Automatic Astronomy,''
     ed. G. W. Henry \& J. A. Eaton (San Francisco: ASP), 86

\refitem
Treffers, R. R., Richmond, M. W., \& Filippenko, A. V. 1992, in
     ``Robotic Telescopes in the 1990s,'' ed. A. V. Filippenko (San
     Francisco: ASP), 115

\refitem
Tripp, R., \& Branch, D. 1999, ApJ, 525, 209

\refitem
Tsvetkov, D. Yu., \& Pavlyuk, N. N. 1997, Astron. Let., 23, 26

\refitem
Udalski, A., Szymanski, M., Kubiak, M., Pietrzynski, G., Soszynski, I.,
   Wozniak, P., \& Zebrun, K. 1999, Acta Astron., 49, 223

\refitem
Van Dyk, S. D., Treffers, R. R., Richmond, M. W., Filippenko, A. V.
   \& Paik, Y. B. 1994, IAU Circ. No. 6105

\refitem
Whitmore, B., Heyer, I., \& Casertano, S. 1999, PASP, 111, 1559

\refitem
Zaritsky, D., Kennicutt, R. C., Jr., \& Huchra, J. P. 1994, ApJ, 420, 87

\refitem
Zehavi, I., Riess, A. G., Kirshner, R. P., \& Dekel, A. 1998, ApJ, 503, 483

\begin{figure}[ht]
\vspace*{140mm}
\includegraphics{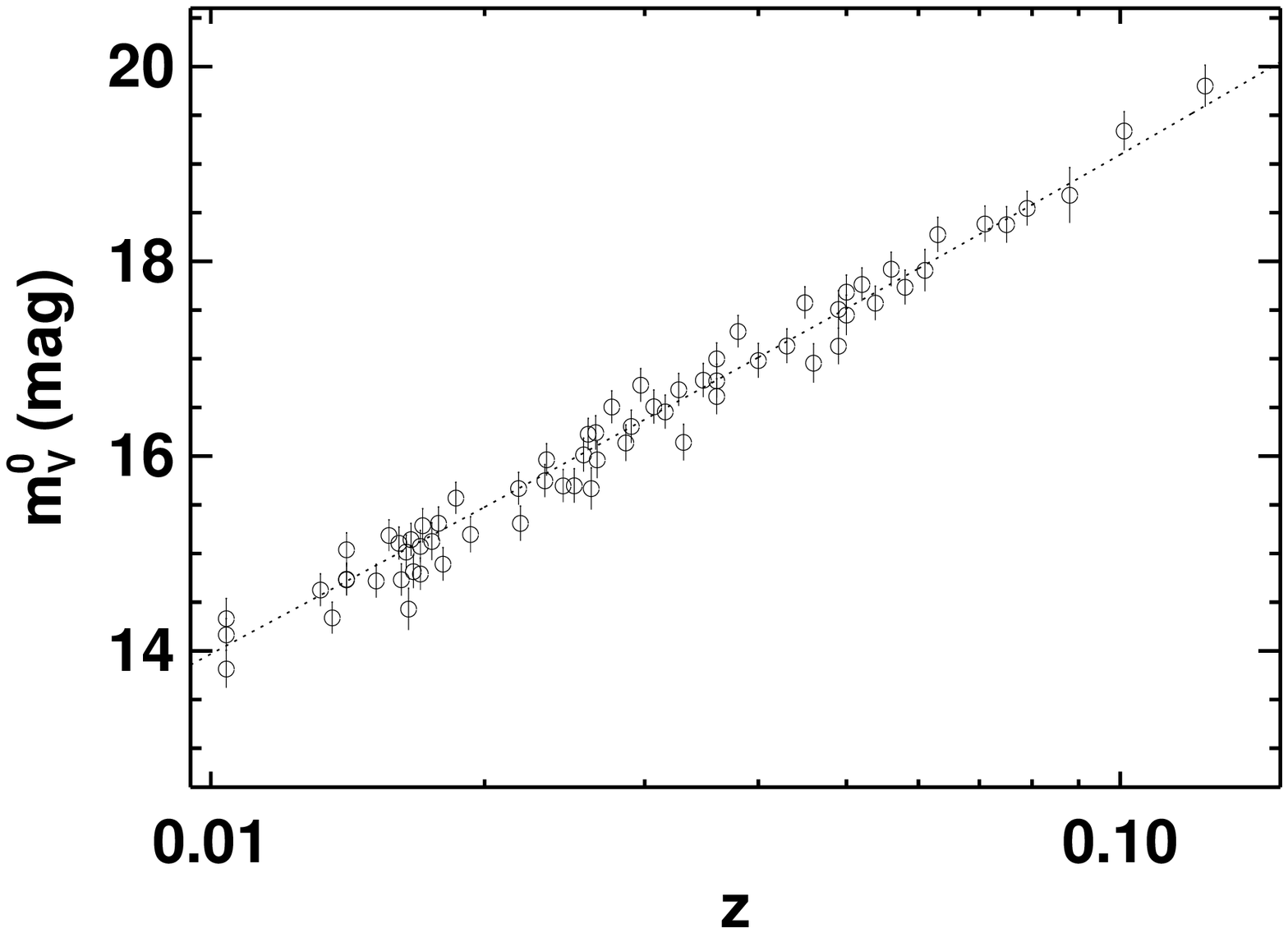}
\caption { }
\end{figure}

\begin{figure}[ht]
\vspace*{100mm}
\caption {Figure 2.  See attached jpeg image }
\end{figure}

\begin{figure}[ht]
\vspace*{100mm}
\caption {Figure 3.  See attached jpeg image }
\end{figure}

\begin{figure}[ht]
\vspace*{150mm}
\includegraphics{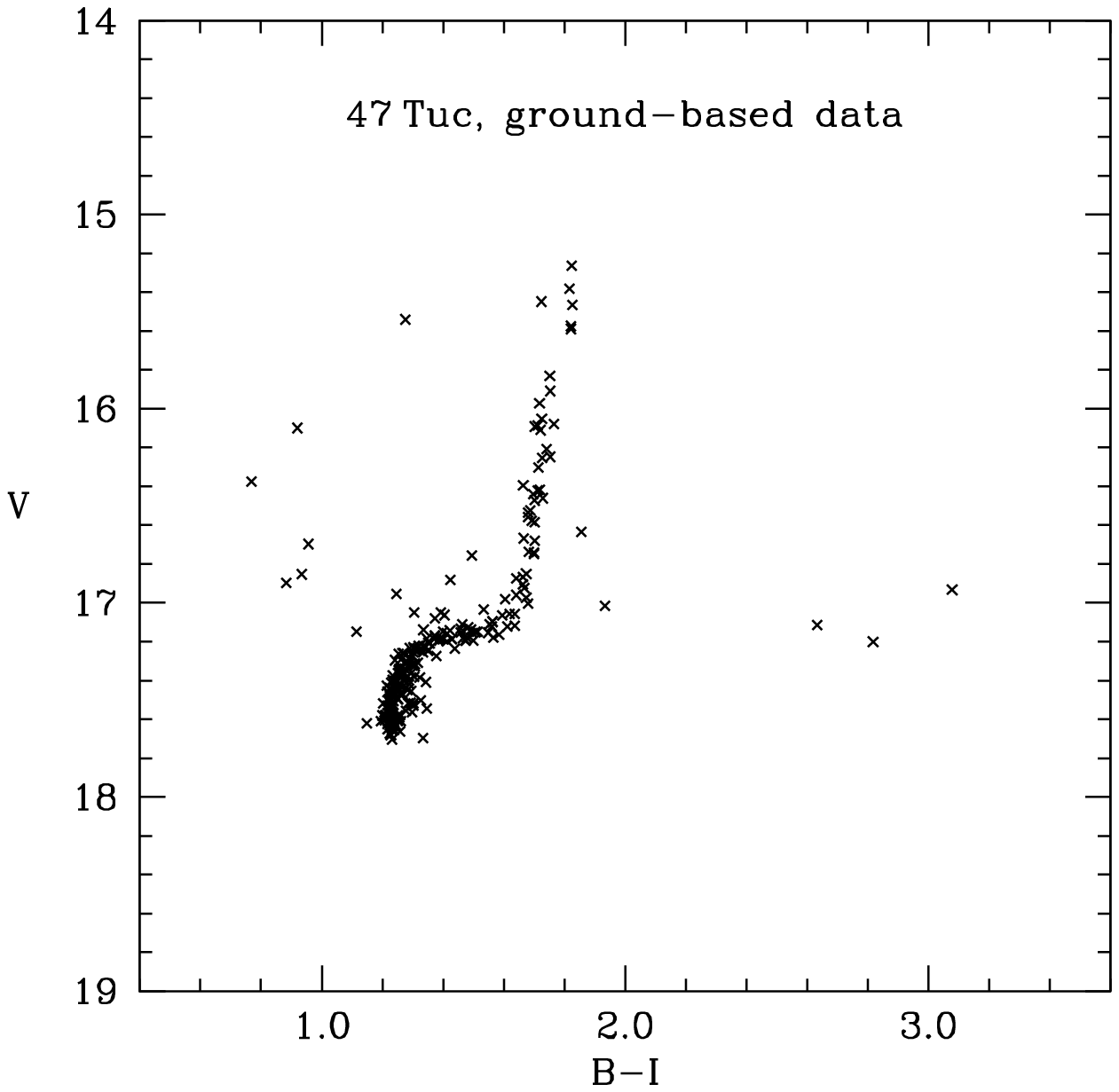}
\caption { }
\end{figure}

\begin{figure}[ht]
\vspace*{70mm}
\figurenum{5}
\includegraphics{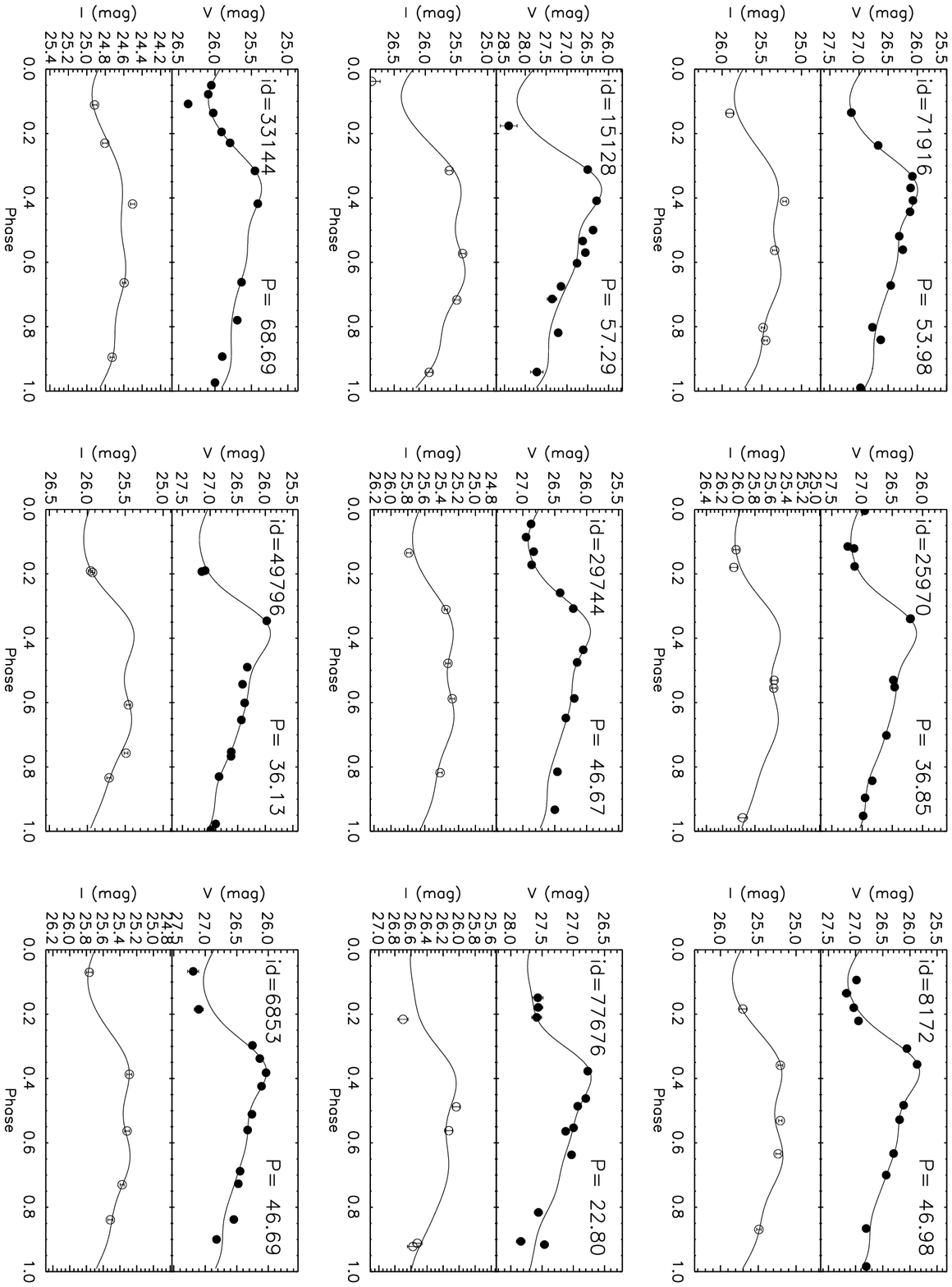}
\caption { }
\end{figure}

\begin{figure}[ht]
\vspace*{70mm}
\figurenum{5}
\includegraphics{f5b.ps}
\caption { }
\end{figure}

\begin{figure}[ht]
\vspace*{70mm}
\figurenum{5}
\caption {Figure 5c-5h contained in online ApJ }
\end{figure}

\vfill
\eject

\begin{figure}[ht]
\vspace*{100mm}
\figurenum{6}
\caption {Figure 6  See attached jpeg image }
\end{figure}

\begin{figure}[ht]
\vspace*{140mm}
\figurenum{7}
\includegraphics{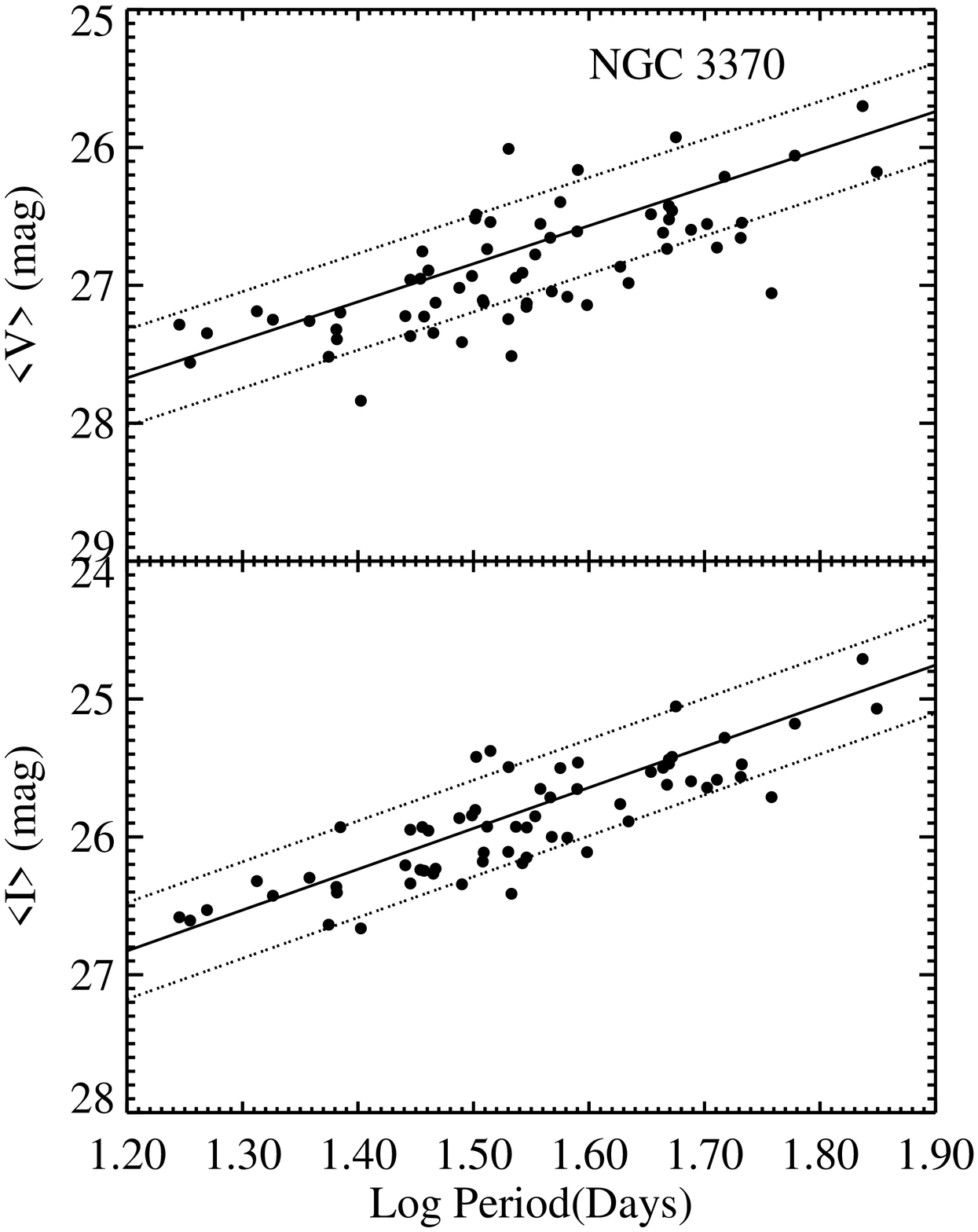}
\caption { }
\end{figure}

\begin{figure}[ht]
\vspace*{140mm}
\figurenum{8}
\includegraphics{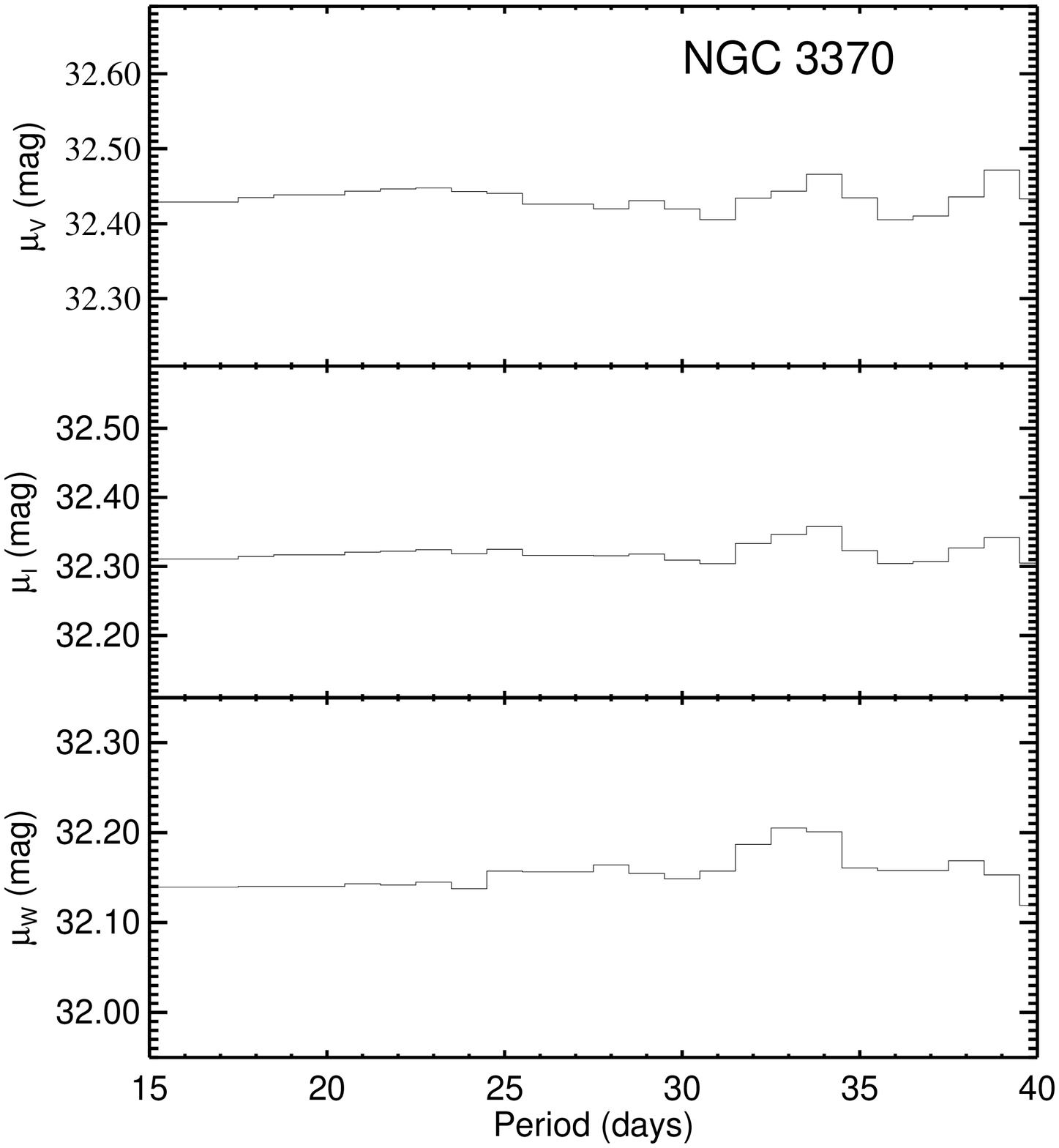}
\caption { }
\end{figure}

\vfill
\eject

\begin{figure}[ht]
\vspace*{140mm}
\figurenum{9}
\includegraphics{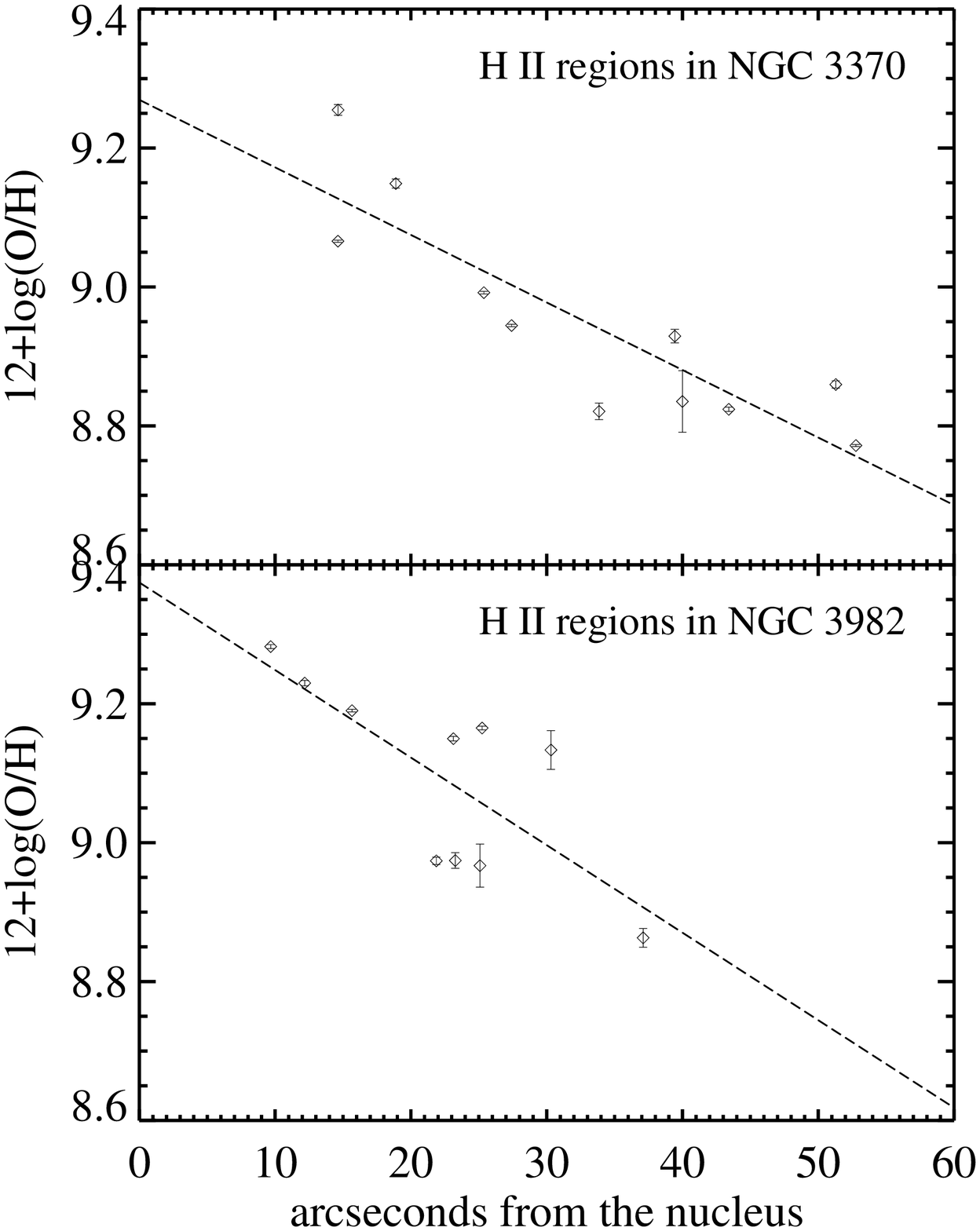}
\caption { }
\end{figure}

\begin{figure}[ht]
\vspace*{120mm}
\figurenum{10}
\caption {Figure 10  See attached jpeg image }
\end{figure}

\begin{figure}[ht]
\vspace*{100mm}
\figurenum{11}
\caption {Figure 11  See attached jpeg image }
\end{figure}

\begin{figure}[ht]
\vspace*{90mm}
\figurenum{12}
\includegraphics{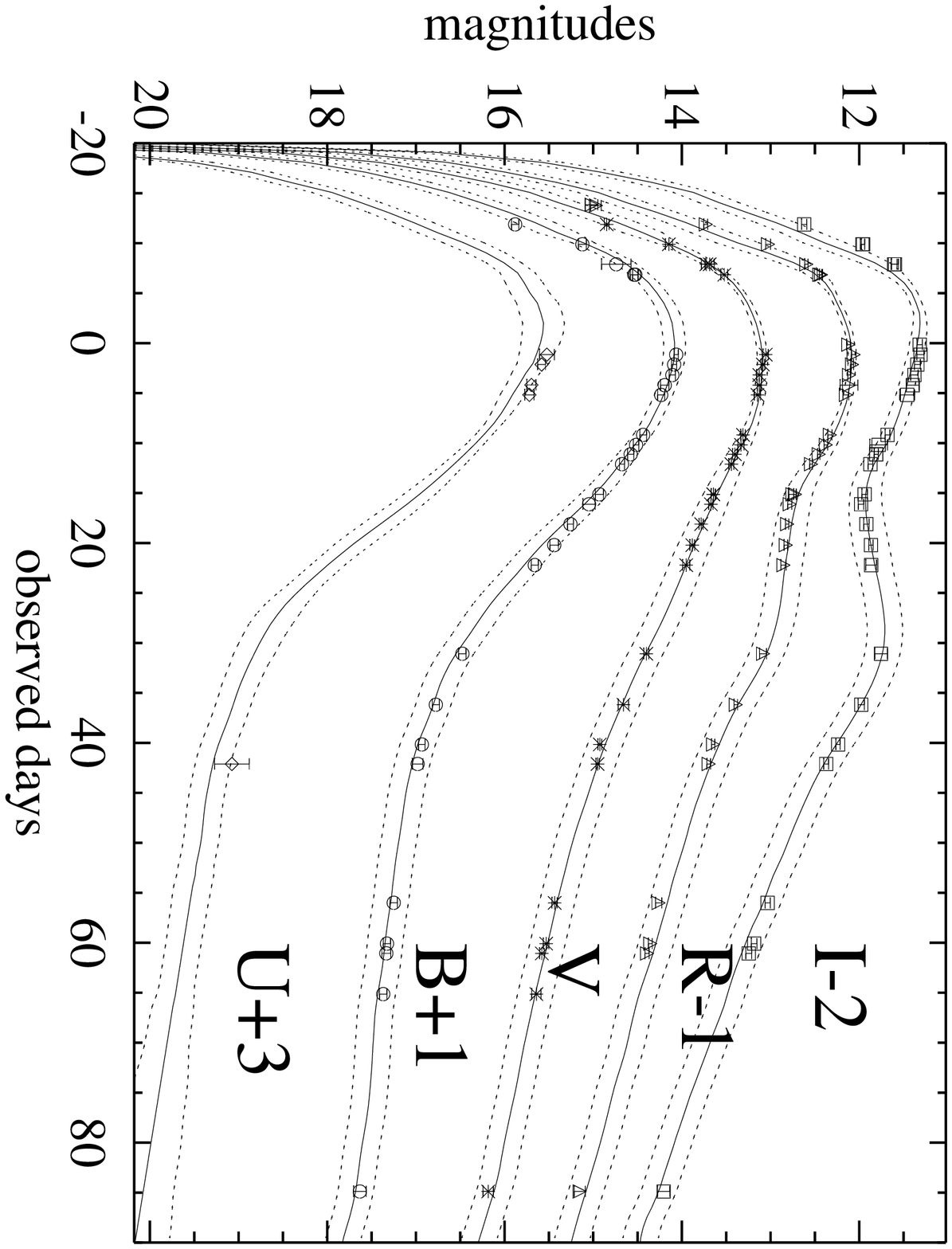}
\includegraphics{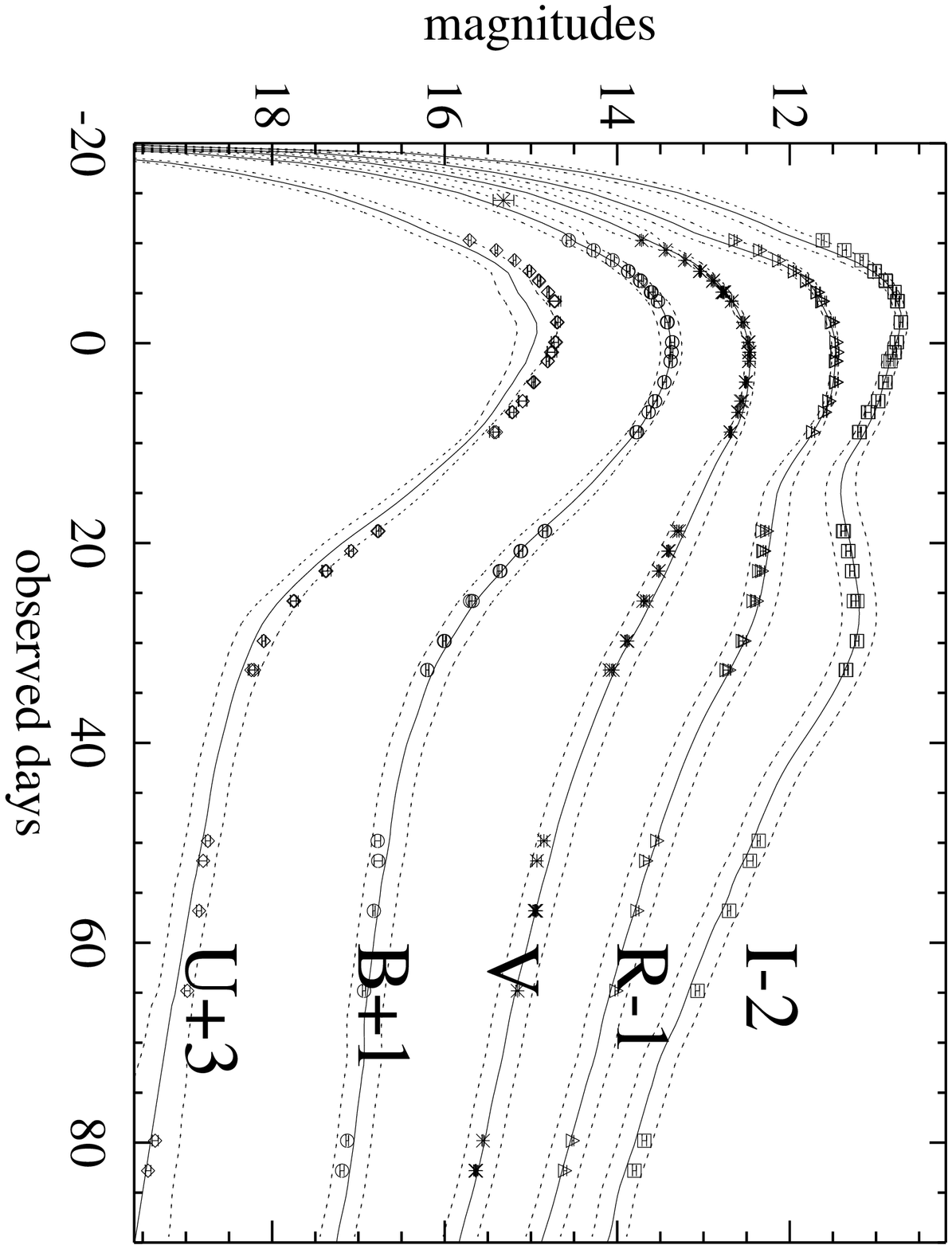}
\caption { }
\end{figure}

\end{document}